\documentclass[a4paper,fleqn,usenatbib]{mnras}

\usepackage[T1]{fontenc}
\usepackage{ae,aecompl}

\usepackage{graphicx}	
\usepackage{amsmath}	
\usepackage{amssymb}	
\usepackage{times}
\usepackage{xcolor}
\usepackage{newtxtext,newtxmath}
\hypersetup{draft} 
\usepackage[normalem]{ulem}
\usepackage{newtxtext,newtxmath}

\def\Rvir{R_{\rm vir}}
\def\R200c{r_{\rm 200c}}
\def\Mvir{M_{\rm vir}}
\def\Lstar{L_\star}
\def\tff{t_{\rm ff}}
\def\tcool{t_{\rm cool}}
\def\tctff{\tcool/\tff}
\def\Kbase{K_{\mathrm{e,base}}}
\def\Kpre{K_{\mathrm{e,pre}}}
\def\Ke{K_{\mathrm{e}}}
\def\nH{n_{\rm H}}
\def\Msun{\mathrm{M}_{\odot}}
\def\mp{m_{\mathrm{p}}}
\def\Klow{K_{\rm low}}

\definecolor{simb}{rgb}{0.477504, 0.821444, 0.318195}
\definecolor{simc}{rgb}{0.157729 0.485932 0.558013}
\definecolor{simr}{rgb}{0.187231 0.414746 0.556547}
\definecolor{simw}{rgb}{0.127568 0.566949 0.550556}
\definecolor{simz}{rgb}{0.1941   0.399323 0.555565}

\title[Thermal Instability in the $\Lstar$ CGM]{Thermal Instability in the CGM of $\Lstar$ Galaxies: Testing ``Precipitation'' Models with the FIRE Simulations}

\author[C. J. Esmerian et al.]{
\parbox{\textwidth}{
Clarke J. Esmerian,$^{1,2}$\thanks{E-mail: cesmerian@uchicago.edu}
Andrey V. Kravtsov,$^{1,2,3}$
Zachary Hafen,$^{4,5}$
Claude-Andr\'e Faucher-Gigu\`ere,$^4$
Eliot Quataert,$^6$
Jonathan Stern,$^4$
Du\v{s}an Kere\v{s},$^7$
Andrew Wetzel$^8$}
\vspace{0.4cm} \\
$^1$Department of Astronomy \& Astrophysics, The University of Chicago,
Chicago, IL 60637 USA\\
$^2$Kavli Institute for Cosmological Physics, The University of Chicago,
Chicago, IL 60637 USA\\
$^3$Enrico Fermi Institute, The University of Chicago, Chicago, IL 60637\\
$^4$Department of Physics and Astronomy and Center for Interdisciplinary Exploration and Research in Astrophysics (CIERA),\\
Northwestern University, 2145 Sheridan Road, Evanston, IL 60208, USA\\
$^5$ Center for Cosmology, Department of Physics and Astronomy, University of California, Irvine, CA 92697, USA\\
$^6$Department of Astronomy and Theoretical Astrophysics Center, University of California Berkeley, Berkeley, CA 94720 \\
$^7$Department of Physics, Center for Astrophysics and Space Sciences, University of California, San Diego, 9500 Gilman Drive, La Jolla, CA 92093, USA \\
$^8$Department of Physics, University of California, Davis, CA 95616, USA\\
}

\date{Accepted XXX. Received YYY; in original form ZZZ}

\pubyear{2020}

\begin{document}
\label{firstpage}
\pagerange{\pageref{firstpage}--\pageref{lastpage}}
\maketitle

\begin{abstract}
    We examine the thermodynamic state and cooling of the low-$z$ Circum-Galactic Medium (CGM) in five FIRE-2 galaxy formation simulations of Milky Way-mass galaxies. We find that the CGM in these simulations is generally multiphase and dynamic, with a wide spectrum of largely nonlinear density perturbations sourced by the accretion of gas from the Inter-Galactic Medium (IGM) and outflows from both the central and satellite galaxies. We investigate the origin of the multiphase structure of the CGM with a particle tracking analysis and find that most of the low entropy gas has cooled from the hot halo as a result of thermal instability triggered by these perturbations. The ratio of cooling to free-fall timescales $\tctff$ in the hot component of the CGM spans a wide range $\sim 1-100$ at a given radius, but exhibits approximately constant median values $\sim 5-20$ at all radii $0.1 \Rvir < r < \Rvir$. These are similar to the $\approx 10-20$ value typically adopted as the thermal instability threshold in ``precipitation'' models of the ICM. Consequently, a one-dimensional model based on the assumption of a constant $\tctff$ and hydrostatic equilibrium approximately reproduces the number density and entropy profiles of each simulation, but only if it assumes the metallicity profile and temperature boundary condition taken directly from the simulation. We explicitly show that the $\tctff$ value of a gas parcel in the hot component of the CGM does not predict its probability of subsequently accreting onto the central galaxy. This suggests that the value of $\tctff$ is a poor predictor of thermal stability in gaseous halos in which large-amplitude density perturbations are prevalent.
\end{abstract}




\section{Introduction}

\hspace{\parindent} A physical understanding of the Circum-Galactic Medium (CGM) remains one of the key challenges of galaxy formation theory \citep[see][for a recent review]{Tumlinson2017}. This is because both the accretion of gas from the Inter-Galactic Medium (IGM) and outflows driven by supernovae and Active Galactic Nuclei (AGN) are expected to leave imprints on its thermodynamic properties \citep[e.g.,][]{Sharma2018}, and the physics governing both of these complicated processes remains deeply uncertain. 

One of the key results of observational studies of the CGM around galaxies is that ``cool'' $\sim 10^4$~K gas is often found to co-exist with higher-temperature $T\sim 10^5-10^6$ K gas \citep[e.g.,][]{Chen2000, Thom2008, Tripp2008, Werk2013, Werk2016, Ng2019, Berg2019}. Around our own galaxy and our nearest neighbor M31, there is evidence for a similar thermodynamic structure in the form of ``high-velocity'' clouds of neutral hydrogen \citep[e.g.][and references therein]{Putman2012}. Signatures of this cool gas are even observed in the much hotter ($\sim 10^7-10^8$ K) intracluster medium (ICM) of $\approx 50-70\%$ of the nearby massive groups  \citep[e.g.,][]{Gauthier2009, Gauthier2010, Gauthier2011, Huang2016, OSullivan2017, Chen2018, Zahedy2019} and clusters \citep[e.g.,][]{Cavagnolo2009,Hogan2017}. The ubiquity of the {\it multiphase} nature of the CGM and ICM is not yet understood. Although models and simulations do produce gas with a wide range of temperatures in gaseous haloes of galaxies and clusters, the observed column density distribution and covering fraction of cool gas is generally not fully reproduced \citep[see, e.g.][]{Liang2016, OppenheimerSchaye2018, Lehner2020}. 

Thermal instability driven cooling \citep{Parker1953,Weymann1960,Field1965} is one possible explanation for the multiphase structure and prevalence of cold gas in galactic haloes \citep[e.g.,][]{MoMiraldaEscude1996, Maller2004}. In the context of galaxy groups and clusters, models based on numerical simulations carried out under the assumption of global thermal equilibrium have proven to be quite successful in reproducing the structure of hot ICM. The assumption of global thermal balance, whereby heating from central supermassive black hole feedback is assumed to offset cooling losses of the hot ICM \citep[e.g.,][and references therein]{McNamara2007,Wang2019}, is motivated by observations that show a general lack of global cooling in the cores of groups and clusters, despite short cooling times of hot gas \citep[see, e.g.,][for a review]{Peterson2006}. 

Specifically, \citet{McCourt2012} and \citet{Sharma2012} used idealized simulations of hot stratified atmospheres in global thermal balance to argue that the onset of thermal instability in such environments is controlled by the ratio of cooling and free-fall timescales, $\tctff$, where
\begin{equation}
    \tcool = \frac{e}{de/dt} = \frac{nk_{\rm B} T/(\gamma - 1)}{\mathcal{C}_{\rm rad} - \mathcal{H}_{\rm rad}},
\end{equation}
\begin{equation}
    \tff = \sqrt{\frac{2r}{g}}
\end{equation}
and $e$ is the thermal energy density, $n = \rho/(\mu\mp)$ the total free particle number density, $k_{\rm B}$ the Boltzmann constant, $T$ the temperature, $\gamma = 5/3$ the adiabatic index for an ionized plasma, $\mathcal{C}_{\rm rad}$ and $\mathcal{H}_{\rm rad}$ the respective volumetric radiative cooling and heating rates, $r$ the radial distance from halo centre, and $g$ the gravitational acceleration at that radius. Their results suggested the existence of a threshold value of $\tctff\approx 10$ below which isobaric density perturbations are thermally unstable and ``precipitate'' out of the hot ICM, producing a multiphase medium. 

\citet{McCourt2012} and \citet{Sharma2012} proposed a scenario in which the ICM is maintained in a state near the threshold value of $\tctff\approx 10$ by repeated cycle of thermal instability driven cooling and heating by the central AGN in response. Such a scenario is qualitatively supported by observations  \citep[e.g.,][]{Salome2006,Tremblay2016,Babyk2018,Lakhchaura2018,Olivares2019} and simulations \citep{Gaspari2012,Li.Bryan2014a,Li.Bryan2014,Li2015,Meece2015,Wang2019,Beckmann2019} of the hot ICM in galaxy clusters. 

Analytic models based on the assumption that ICM gas is maintained near a constant $\tctff\approx 10$ threshold appear to successfully reproduce the shape of radial profiles of ICM density and entropy in the central regions of many groups and clusters \citep[][]{Voit.Donahue2015,Voit2015,Voit2018a}, although recent analyses indicate that the ICM in the cores of many clusters spans a fairly wide range of values $\tctff\sim 10-25$ \citep{Hogan2017,Pulido2018}. Moreover, the amount of cold molecular gas and strength of AGN activity does not show correlation with $\tctff$ \citep{Pulido2018}. A number of theoretical studies have used simulations to explore the triggering of thermal instabilities in environments with a wider range of properties, as well as exploring more sophisticated scenarios and processes \citep{Gaspari2013,Meece2015,Singh.Sharma2015,Choudhury.Sharma2016,Voit2017,Prasad2018,Choudhury2019}. In parallel, the possible effects of nonlinear perturbations in the CGM generated by uplift of gas due to outflows and turbulence have been considered with analytical modeling \citep{Voit2017,Voit2018b,Voit2019a}. These studies found that precipitation of cool gas can occur when $\tctff\sim 10-20$, especially when seed gas density perturbations produced by gas accretion, AGN feedback or tidal interactions are large \citep[e.g.,][]{Pizzolato.Soker2005,Joung2012}.   

As the exploration of and debate on the physics of ICM cooling in groups and clusters continues, some recent investigations have extended the precipitation scenario to galaxy scales and argued that circum-galactic thermal instability may play a central role in the cycle of gas accretion, star formation, and feedback \citep{Soker2010,Sharma2012b,Voit2015,Voit2018a,Voit2019b,Voit2019a}. Although the overall dynamics of gravitational collapse is approximately self-similar in the $\Lambda$CDM cosmology \citep[e.g.,][]{Kravtsov.Borgani2012}, the interplay of gas shock heating and cooling during the formation of galaxy-sized systems is qualitatively different from that of cluster-scale haloes. 

In particular, much of the gas in these systems can accrete in cold streams without substantial shock heating \citep{Keres2005,Keres2009,Dekel.Birnboim2006,Ocvirk2008,Dekel2009,FaucherGiguere2011, FaucherGiguere2015, Rosdahl2012} and a hot, diffuse gaseous CGM can be maintained only in haloes of total mass $\gtrsim 0.5-1\times 10^{12}\, \Msun$ \citep{Birnboim.Dekel2003,Dekel.Birnboim2006,FaucherGiguere2011b,Fielding2017,Stern2020}. In haloes at or below this threshold mass, a significant fraction of gas is expected to cool onto central galaxies, leaving a gaseous halo with a nearly-constant entropy core \citep[e.g.,][see also Figs.~\ref{fig:radpro} and \ref{fig:oneD_models} below]{Sharma2012b}. This shallow entropy distribution is expected to be conducive to runaway global cooling, as expected theoretically in the regime when $\tctff\lesssim 1$ \citep{Singh.Sharma2015,Choudhury2019}. 

This cooling, together with feedback-driven gas flows and turbulence, can result in large deviations from hydrostatic equilibrium \citep[e.g.][]{Oppenheimer2018}.
A large fraction of CGM is expected to be well mixed, with some gas re-accreted onto the galaxies after being expelled in galactic outflows \citep[e.g.,][]{AnglesAlcazar2017,Muratov2017,Hafen2019,Hafen2019b,Borrow2019}. At the same time, the outflows in Milky Way-mass galaxies are expected to subside or cease completely at low redshifts \citep[][Stern et al. in prep]{Muratov2015}, and it is not clear whether galaxies can maintain global thermal balance in their CGM at any stage of their evolution. 

The differences between the evolution of galaxy-scale gaseous haloes and those of group- and cluster-scale systems are thus expected to be substantial. The applicability of the precipitation scenario in this regime and the overall role of thermal instability on galactic scales are thus not yet understood. Although some early simulations of idealized, equilibrium galactic CGM predicted the continuous formation of cold clouds \citep{Connors2006,Kaufmann2009}, later studies have not confirmed these results, but attributed this to numerical effects \citep{Nelson2013,Huang2019}. At the same time, a systematic test of models based on the precipitation scenario with cosmological simulations of galaxy formation has not yet been done. 

\begin{figure*}
\centering
\includegraphics[height=0.9\textheight]{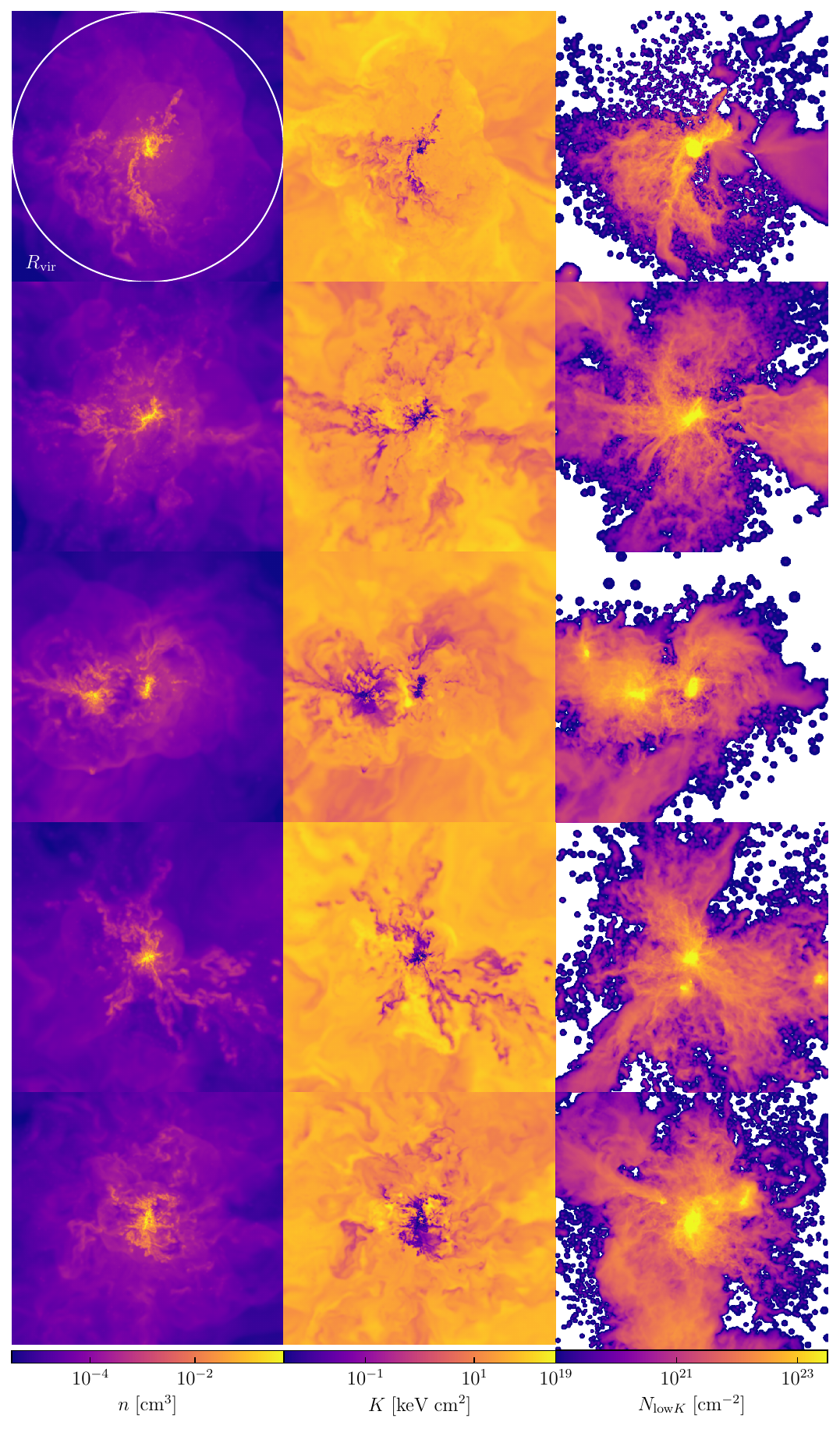}
\caption{Maps of the simulated CGM. Each row shows one of the simulations in our sample (from top to bottom: {\bf m12b, m12c, m12r, m12w, m12z}) at $z=0.25$. Left and middle panels show gas number density of all free particles (including electrons) and specific entropy, respectively, on a random two-dimensional slice through the CGM of these simulations.  The rightmost panel shows the column density of low-entropy ($K < \Klow \equiv  5 \;{\rm keV}\,{\rm cm}^2$) gas projected along the same line-of-sight within $\pm \Rvir$ of the halo center, calculated using the same number density as shown on the leftmost panel. Panels are $2R_{\rm vir} \approx 400\, {\rm kpc}$ on each side. All galaxies in these simulations display a dynamic, multiphase CGM consisting of a high-entropy hot halo with embedded low-entropy regions. These maps were generated with the publicly available \texttt{meshoid} package~\citep{Meshoid}.}
\label{fig:maps}
\end{figure*}

This is precisely the goal of our paper. To this end, we use a suite of five high-resolution zoom-in cosmological simulations of Milky Way-mass galaxies from the FIRE-2 project to analyze the thermodynamic properties of the CGM at $z\approx 0$ and their physical origin. Although the physics of galaxy formation is still being actively debated, the simulations we use in this study have been shown to reproduce a number of key properties such as morphologies, sizes, and metallicities of observed galaxies with similar luminosities reasonably well \citep[see, e.g.,][]{FIRE2}. It is thus interesting to examine whether these properties have been shaped by the thermal instability-driven cooling-feedback cycle envisioned in the precipitation scenario. Specifically, we compare the predictions of models motivated by precipitation to the distributions of gas density, entropy, and $\tctff$ in the CGM of these simulations (Sections \ref{sec:genprop} and \ref{sec:modelcomp}). We also investigate the physical origins and fate of low-entropy gas in these haloes by tracking the thermodynamic evolution of gas tracers -- Lagrangian cells that move with the gas flow and track the evolution of its thermodynamic variables  (Sections \ref{sec:lowKorigins} and \ref{sec:accretion}). 

We find that the CGM in these simulations shows a wide dispersion in all gas dynamical quantities at any given radius, indicating a multiphase thermodynamic structure and significant deviations from hydrostatic equilibrium (HSE). Nevertheless, since the median $\tctff$ value in the hot phase is roughly constant at $\sim10$, and HSE deviations are not significantly larger than $\sim50\%$\footnote{as quantified by the ratio of thermal pressure and gravitational potential gradients, see Section~\ref{sec:genprop}}, one-dimensional ``precipitation'' models constructed assuming $\tctff=10$ and HSE \citep[e.g.,][]{Sharma2012} can approximately match the median density and entropy profiles in the simulations, provided they are calculated using the metallicity profile and temperature at the virial radius (which sets the boundary condition needed to integrate the HSE equation) directly from the simulations. 

Moreover, we \textit{do} find evidence that a significant fraction of the low-entropy gas in these haloes results from thermal instability. These instabilities are seeded mostly by large ($\delta\rho/\rho \equiv \delta \gtrsim 1$), compressive density perturbations resulting from cosmological accretion flows, feedback-driven outflows, and gas tidally or ram-pressure stripped from the central and satellite galaxies. We find that although the median $\tctff$ of the hot halo gas is roughly constant, $\tctff$, value of $\tctff\sim 5-20$ at $0.1\lesssim r/R_{\rm vir}\lesssim 1$, there is a wide distribution of the local values of this ratio at a given radius. Moreover, the value of $\tctff$ for a given parcel of gas in the hot phase of the CGM does not strongly predict its likelihood of undergoing thermal instability and subsequently accreting onto the galaxy. 

\section{Simulations and analysis methods}
\label{sec:simulations}

\subsection{FIRE-2 ``zoom-in'' galaxy formation simulations}
\label{sec:fire2sims}

Our analysis uses simulations of galaxies forming in $\approx 10^{12}\, \Msun$ haloes from the Feedback in Realistic Environments (FIRE)\footnote{\url{http://fire.northwestern.edu}} project with the  ``FIRE-2'' model of galaxy formation \citep{FIRE2}. Interested readers should refer to that paper for details, but here we summarize the main features of the model. 

The simulations use the Meshless-Finite-Mass method of the GIZMO \citep{Hopkins2015GIZMO}\footnote{\url{http://www.tapir.caltech.edu/~phopkins/Site/GIZMO.html}} gravity plus hydrodynamics code to solve the equations of inviscid gas dynamics coupled to the gravitational evolution of collisionless stars and cold dark matter. Source terms from radiative cooling and heating due to photo-ionization and recombination, Compton, free-free, photoelectric and dust, collisional, cosmic ray,\footnote{including {\it only} effective heating terms in the ISM, since the simulations we analyze here do not evolve a cosmic ray fluid component} molecular, metal-line and fine-structure processes are included for gas with temperatures from $10-10^{10}$K using tabulated {\sc Cloudy} \citep{Ferland1998} calculations. These calculations consider contributions from each of the 11 tracked chemical species. Modifications due to both local radiation sources and a cosmological UV background \citep{FaucherGiguere2009}, accounting for self-shielding, are included. Star formation is chosen to occur in self-gravitating \citep{Hopkins2013sfr_criteria}, molecular \citep{KrumholzGnedin2011}, Jeans unstable, and sufficiently dense ($\nH > 1000\;{\rm cm}^{-3}$) gas cells with 100$\%$ efficiency (in molecular gas mass) per free-fall time. Stellar feedback processes are tabulated from stellar evolution models ({\sc Starburst99}; \citealt{Leitherer1999}) by treating individual star particles as single age and metallicity stellar populations with the \citet{Kroupa2001} IMF. This feedback prescription includes the energy, momentum, mass, and metals due to supernovae (Type Ia and II) and stellar mass loss (OB $\&$ AGB), radiation pressure, and photo-ionization and photoelectric heating.

\begin{table*}
\caption{Properties of the five haloes and the CGM they host in the FIRE-2 simulations at $z=0.25$ used in this study. $\Mvir$: halo virial mass. $\Rvir$: halo virial radius. $f_b/f_{b,\rm cosmo}$: baryon fraction within $\Rvir$ as a fraction of the cosmic budget. $M_*$: central galaxy stellar mass. $M_{\rm CGM}$: total gas mass of the CGM. $f_{{\rm low}\;K}$: fraction of CGM gas mass at low-entropy $K < \Klow\equiv 5\;{\rm keV}\;{\rm cm}^2$. $f_{{\rm low}\;K,{\rm cooled}}$: fraction of low-entropy CGM gas mass that was previously at high entropy $K > \Klow$ at {\it any} earlier cosmological time. $t_{\rm cool}$, $\tctff$: median value of the cooling timescale and cooling-to-free-fall timescale ratio in the high-entropy gas, reported at $0.2\Rvir$ and $\Rvir$. Simulations ${\bf m12b, m12c, m12z}$ were introduced in \citet{Garrison-Kimmel2019}, ${\bf m12r, m12w}$ in \citet{Samuel2020}.}                                                
\resizebox{\textwidth}{!}{\begin{tabular}{cccccccccccccccc}
\hline
\hline
Name    & $M_{\rm vir}$ & $R_{\rm vir}$ & $f_b/f_{b,\rm cosmo}$ & $M_{*}$   & $M_{\rm CGM}$ & $f_{{\rm low}\;K}$    & $f_{{\rm low}\;K,{\rm c}}$    & $(t_{\rm cool})_{0.2R_{\rm vir}}$  & $(t_{\rm cool})_{R_{\rm vir}}$ & $\left( \frac{t_{\rm cool}}{t_{\rm ff}}\right)_{0.2R_{\rm vir}}$   & $\left(\frac{t_{\rm cool}}{t_{\rm ff}}\right)_{R_{\rm vir}}$  & \\
    & $[10^{12} M_{\odot}]$ & $[{\rm kpc}]$ &   & $[10^{10} M_{\odot}]$ & $[10^{10} M_{\odot}]$ &   &   & [Gyr] & [Gyr] &   &   & \\
\hline
{\bf m12b}  & 1.27  & 245   & 0.73  & 7.76  & 5.23  & 0.18  & 0.87  & 7.3 & 48.6 & 21.2 & 12.9\\
{\bf m12c}  & 0.79  & 210   & 0.74  & 4.80  & 3.06  & 0.23  & 0.88  & 3.8 & 25.9 & 11.3 & 7.0\\
{\bf m12r}  & 0.69  & 200   & 0.52  & 1.39  & 3.44  & 0.46  & 0.83  & 4.7 & 25.4 & 11.2 & 6.0\\
{\bf m12w}  & 0.89  & 218   & 0.64  & 4.56  & 2.80  & 0.38  & 0.80  & 3.5 & 21.8 & 9.6  & 12.2\\
{\bf m12z}  & 0.67  & 198   & 0.62  & 1.41  & 4.02  & 0.49  & 0.73  & 2.3 & 13.5 & 5.9  & 4.4\\
\label{table:simulations}
\end{tabular}}
\end{table*}

We analyze five zoom-in cosmological galaxy formation simulations that focus resolution on the Lagrangian regions of $\sim 10^{12} \Msun$ dark matter haloes at $z=0$. The names and basic properties of the host haloes and their CGM are summarized in the first four columns of Table \ref{table:simulations}. The simulations adopt flat $\Lambda$CDM cosmology with  $H_0 = 70 {\rm km}\;{\rm s}^{-1}{\rm Mpc}^{-1}$, $\Omega_{\rm m0} = 0.27$, $\Omega_{\Lambda} = 1 - \Omega_{\rm M} = 0.73$, and $\Omega_{\rm b} = 0.049$. Four of our simulations ({\bf m12b, m12c, m12r, m12w}) have a gas element mass resolution of $m_{\rm b} = 7070\Msun$, a dark matter particle mass of $m_{\rm dm} = 3.52\times10^4 \Msun$, a minimum (adaptive) gas element force softening length of $\epsilon_{\rm g} = 0.5\;{\rm pc}$, a star particle force softening length of $\epsilon_{\star} = 4.0{\rm pc}$, and a dark matter particle force softening length of $\epsilon_{\rm dm} = 40{\rm pc}$. Simulation {\bf m12z} has $m_{\rm b} = 4170\Msun$, $m_{\rm dm} = 2.14\times10^4 \Msun$, $\epsilon_{\rm g} = 0.4{\rm pc}$, $\epsilon_{\star} = 3.2{\rm pc}$, and  $\epsilon_{\rm dm} = 33{\rm pc}$. Softening units are co-moving for $z>9$, physical thereafter. We note that ${\bf m12r, w}$ were specifically selected for hosting a low-$z$ LMC-mass satellite, which ensures some diversity of accretion history within our sample \citep[see][for further details]{Samuel2020}. We explore the effects of simulation resolution on our results in Section~\ref{sec:resolution}

Lagrangian fluid elements do not exchange mass in the MFM hydrodynamics solver. Therefore, the diffusion of heavy elements due to unresolved turbulence in the ISM, CGM, and IGM can be underestimated and needs to be modelled explicitly. All of the simulations analyzed in this paper include an additional prescription for sub-grid metal diffusion based on the \citet{Smagorinsky1963} model, described and tested in \citet{Hopkins2017_diffusion, FIRE2} and \citet{Escala2018}. While previous investigations have shown that the inclusion of metal diffusion does not qualitatively impact the dynamical properties of galaxies themselves \citep{Su2017, FIRE2}, we focus on simulations with the subgrid diffusion model because they represent a more realistic model for metal mixing in the CGM. Since radiative cooling rates are a function of metallicity, investigations of thermal instability are most appropriately addressed with these simulations.

\subsection{Halo catalogues, galaxy definitions, and substructures}
\label{sec:haloes}

To identify dark matter haloes and galaxies, we use the Amiga Halo Finder \citep[AHF;][]{Gill2004, Knollmann2009}. Halo mass is defined within the radius enclosing an overdensity of $\Delta_{\rm vir}$ relative to the critical density of the universe, with $\Delta_{\rm vir}\approx 97$ for our adopted cosmology at $z=0$ \citep{BryanNorman1998}. 

We adopt exactly the same definitions for central and satellite galaxies as \citet{Hafen2019}: each galaxy is defined as the gas and star particles within $R_{\rm gal} = 4 R_{\star, 0.5}$ of the halo center identified by AHF, where $R_{\star, 0.5}$ is the half mass radius for all star particles within $0.15 R_{\rm vir}$. For the primary galaxy of the zoom-in region the estimate of $R_{\rm gal}$ is averaged over a $\approx 500$ Myr window to minimize variations due to major mergers. Gas particles within this radius are considered part of the central galaxy ISM if they have a baryon number density $\nH > 0.13\;{\rm cm}^{-3}$. Satellite galaxies must have at least 10 star particles.

The CGM of the primary halo is defined as all gas elements with galactocentric radii $R_{\rm CGM, inner} < r < R_{\rm vir}$, where $R_{\rm CGM, inner} = \max(1.2 R_{\rm gal}, 0.1R_{\rm vir})$ to ensure that the ISM of the central galaxy is not considered as the CGM. In the gas tracking classifications that we describe below, the gas within a satellite galaxy's $R_{\rm gal}$ is considered to be associated with that satellite's ISM and only this gas is distinguished from the rest of the CGM gas. However, for the calculation of median thermodynamic profiles as a function of galactocentric radius we additionally exclude all gas cells within the tidal radius of any subhalo identified by AHF. This encompasses a larger fraction of the CGM gas than the gas within $R_{\rm gal}$ of satellite galaxies. We choose a more extended definition for this calculation because subhaloes affect thermodynamic properties of the surrounding CGM well beyond the ISM extent of their galaxies (e.g., satellite galaxies can have their own ``CGM''). Throughout the paper we refer to this gas within tidal radii of subhaloes as \textit{substructure}, to distinguish it from the gas associated with \textit{satellite galaxies} in gas tracking.

\subsection{Gas particle tracking and classification}
\label{sec:tracking}

To probe the physical processes operating in the CGM gas of our simulations we use the ``gas particle tracking'' analysis of \citet{Hafen2019}, which we summarize here. Since cells in the MFM scheme move akin to the gas particles in the Smoothed Particle Hydrodynamics (SPH) method but with hydrodynamic interactions computing using a Riemann solver rather than an SPH solver, we can view these cells as gas particles that can be tracked throughout the evolution. This tracking analysis was performed at $z = 0.25$ in the simulations, corresponding to the cosmic epoch for which many observational studies of the CGM around Milky Way-mass galaxies have acquired data. \footnote{Tracking was also done at $z=2$, but we focus on the simulation predictions at low redshift, since these simulated galaxies lack a hot halo at earlier times \citep[see Figure A1 of][]{Hafen2019}, rendering the  precipitation model irrelevant.} For each simulation, a random subset ($10^5$) of cells located in the CGM at $z=0.25$ were classified into four categories based on their trajectory relative to the central and satellite galaxies:
\begin{itemize}
    \item \textit{IGM accretion}: gas particles that have never been in another galaxy. 
    \item \textit{Satellite ISM}: gas particles that are currently inside a satellite galaxy.
    \item \textit{Satellite Wind}: gas particles that have previously been inside a satellite galaxy.
    \item \textit{Central Wind}: gas particles that have previously been inside the main galaxy.
\end{itemize}

While the term ``wind'' is used for gas tracers that left the ISM of the central and satellite galaxies, we do not distinguish between different physical processes such as tidal stripping, ram pressure stripping, or stellar feedback driven outflows in this classification. Further details about this classification procedure, as well as results on the mass fraction, bulk properties, and trends with halo mass of gas in each classification are presented in \citet{Hafen2019}. We extend this analysis by exploring the thermodynamic history of circumgalactic gas, and how it relates to tracking classification, with the specific aim of understanding the processes governing the phase structure of the CGM.   

\subsection{Radiative Cooling Rates}
\label{sec:radcool}

To calculate the cooling time of gas cells in the simulations we use the radiative cooling rates of photoionized gas as a function of density, temperature, and metallicity (assuming a helium mass fraction of $Y=0.25$ and solar abundance ratios of heavy elements, for computational efficiency) tabulated by \citet{Wiersma2009}, which were calculated with the spectral synthesis code \textsc{Cloudy} \citep{Ferland1998} assuming a uniform and redshift-dependant cosmological UV background from \citet{HaardtMadau2001}. These are parameterized as a cooling function $\Lambda(T, Z, \nH)$ such that the net volumetric radiative cooling rate is $\mathcal{C}_{\rm rad} - \mathcal{H}_{\rm rad} = \nH^2\Lambda(T, Z, \nH)$, where $\nH$ is the number density of all hydrogen atoms and ions. While these are the same tables used in the FIRE-2 source code with which the simulations were run, we note that our calculated cooling rates are not {\it identical} to those calculated during simulation run-time because 1) we are not correcting for variation in the helium mass fraction or non-solar heavy element abundance ratios, 2) the total radiative cooling rate calculated in the FIRE simulations includes many other cooling processes \citep[see Appendix B in][]{FIRE2} and 3) the UV background assumed for calculating H and He cooling rates in the simulation code \citep{FaucherGiguere2009} is different than the one assumed for the heavy element cooling \citet{HaardtMadau2001}, while all the rates used in this paper come from tables calculated assuming \citet[][see \citealt{Wiersma2009}]{HaardtMadau2001}. However, we expect that these differences do not affect the results of our analyses. Difference 1 should not significantly affect the calculated cooling rate since the heavy element abundance patterns in the CGM at the late epochs analyzed here should not differ significantly from solar. Differences 2 and 3 only affect low-temperature cooling ($T \lesssim 10^5{\rm K}$) and thus would not affect conclusions about cooling of higher temperature gas, which is the focus of our analysis. 

\section{Results}
\label{sec:results}

\begin{figure*}
\includegraphics[width=0.9\textwidth]{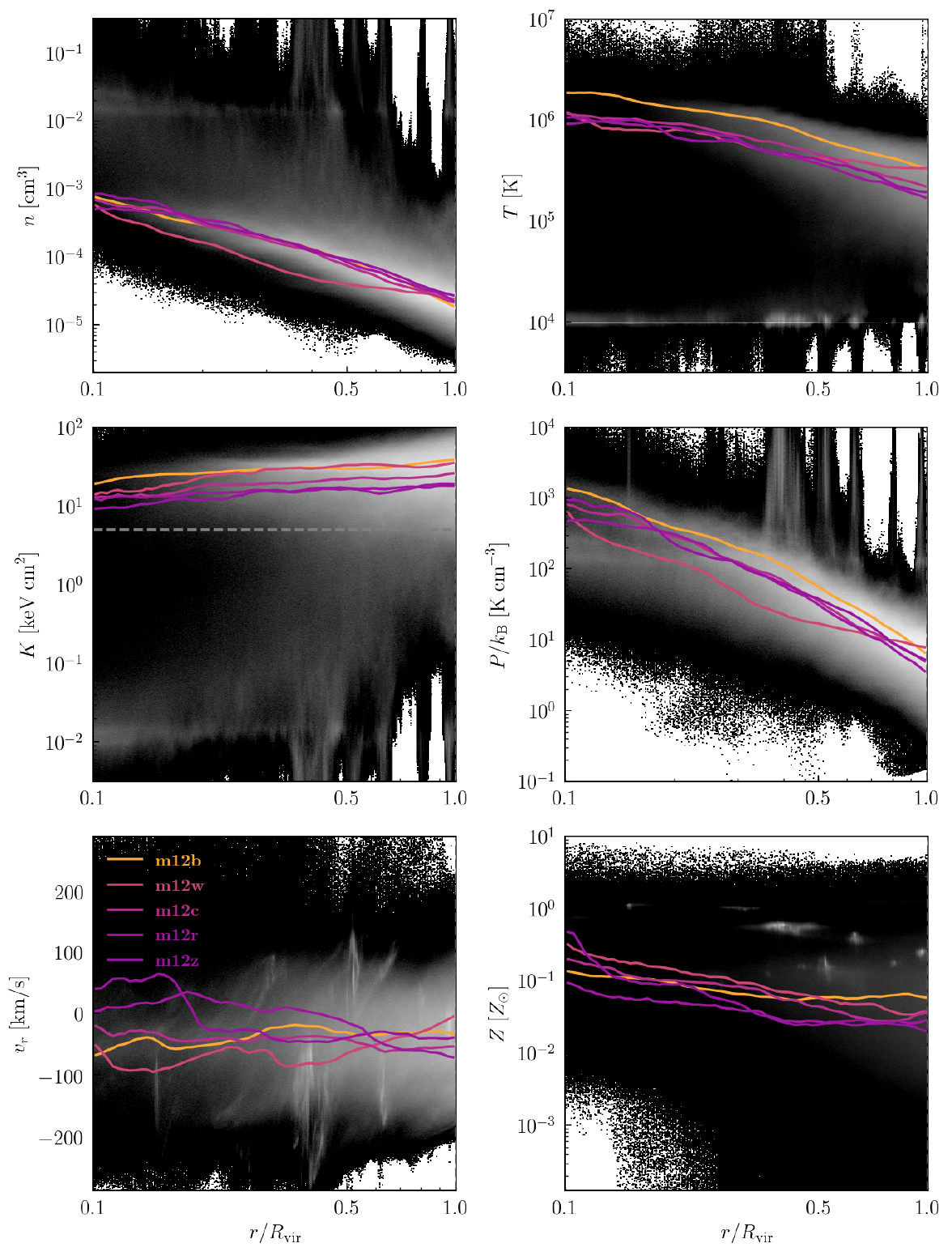}
\caption{Radial profiles. Total free particle number density ($n$), temperature ($T$), specific entropy ($K$), pressure ($P$), radial velocity ($v_r$), and metallicity ($Z$) as a function of galactocentric radius (scaled to the virial radius of each halo) for all five simulations combined at $z=0.25$. Grayscale shading shows the mass-weighted PDF, logarithmically stretched from black (minimum) to white (maximum), of all gas within 0.1 and 1 $R_{\rm vir}$ in all simulations combined. The colored lines are 1D radial profiles for the hot halo in each individual simulation, calculated by taking the median gas element value in each radial bin excluding substructure and low-entropy ($K < \Klow =5\;{\rm keV}\;{\rm cm}^2$) gas. The dashed gray line in the entropy panel indicates this threshold. Simulation profiles are colored from purple to orange (dark to light) by increasing mass (see legend in top left panel and Table~\ref{table:simulations}; we use this simulation-to-color mapping for the remainder of the paper).} 
\label{fig:radpro}
\end{figure*}

\begin{figure}
    \centering
    \includegraphics[width=\columnwidth]{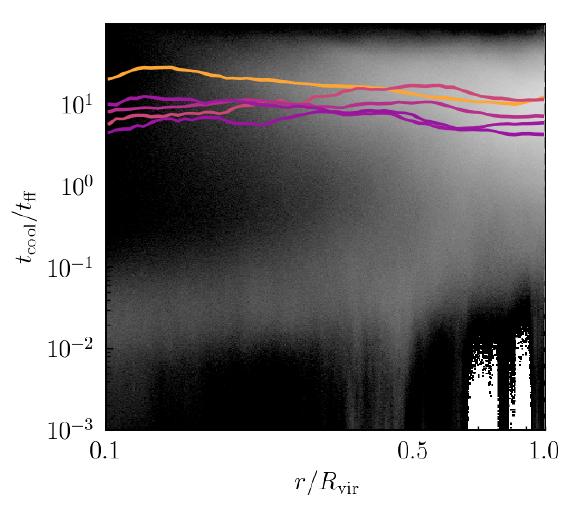}
    \caption{Radial profiles of $\tctff$ for the CGM of the simulated $\Lstar$ galaxies in our suite at $z=0.25$. The grayscale shading and colored lines are the same as in Figure~\ref{fig:radpro}.} 
    \label{fig:prof_tctff}
\end{figure}

\subsection{General properties of low-$z$ $\Lstar$ CGM in the FIRE simulations}
\label{sec:genprop}

We begin with a general overview of the thermodynamic properties of the CGM in the FIRE simulations of galaxies in $\sim10^{12}\,\Msun$ dark matter halos at $z=0.25$. Figure~\ref{fig:maps} shows maps of number density $n$, specific entropy $K$, and column density $N_{\rm col}$ of low-entropy gas in a randomly oriented slice through the $z=0.25$ CGM in three of the galaxies in our sample. $N_{\rm col}$ is calculated along the line-of-sight projected within $\pm \Rvir$ of the galaxy center. Throughout our analysis, we use $n$ to refer to the \textit{total} number density of all particles including electrons. Our analysis focuses on the adiabatic invariant $K = k_{\rm B} T n^{-2/3}$, to which we refer interchangeably as ``entropy'' or ``specific entropy,'' since $K$ does not change under adiabatic compression or expansion and thus directly indicates the phase separation due to cooling that defines thermal instability. We therefore define the ``low-entropy'' CGM to be gas which has $K < \Klow \equiv 5\, \rm{keV}\;\rm{cm}^2$, which we explain with Figure~\ref{fig:K_pdf} in Section~\ref{sec:lowKorigins}. 

The multiphase nature of the CGM in these simulations is readily apparent in Figure~\ref{fig:maps}: diffuse hot gas dominates in both mass and volume, but a significant fraction of the CGM mass at $\lesssim 0.5\Rvir$ of the central galaxy is occupied by clumpy and filamentary regions of relatively low entropy and high-density. This low-entropy gas constitutes $\approx 20-50\%$ of the total CGM mass, of which $\gtrsim 70\%$ has previously cooled from the high-entropy phase (see the 7th and 8th columns in Table~\ref{table:simulations}). A detailed analysis of the physical origins of the low-entropy gas is presented in Section~\ref{sec:lowKorigins}.

Concentric shocks near the virial radius are another striking feature of these maps. They can be seen clearly in the number density map, but are less distinct in the specific entropy slice. The latter reflects the nearly constant entropy in the hot phase of the CGM in these haloes. Any outflows thus have low Mach number and do not significantly heat the gas, which is manifested in relatively small differences in pre- and post-shock specific entropy. Note also that \citet{Muratov2015} and \citet{AnglesAlcazar2017} found that stellar feedback driven outflows in haloes of this mass have smaller mass loading factors or cease entirely by $z\approx 0$ (see also Stern et al. in prep). Both the weakness of these shocks and the small amount of mass (and thus energy) carried by outflows at these redshifts likely result in a total heating rate lower than the net radiative cooling losses of the CGM gas, which would mean these halos are not in global thermal equilibrium.

Additional notable features apparent in the density and entropy maps in Figure~\ref{fig:maps} are the whirl-like fluctuations that are indicative of random motions in the simulated CGM \citep[see e.g.,][for the connection between density fluctuations and turbulence in the ICM of galaxy clusters]{Zhuravleva2014}. We have checked that these motions contribute a typically sub-dominant, but non-negligible contribution to the pressure budget and lead to significant departures from hydrostatic equilibrium ($\sim20-50\%$, as a fraction of the pressure gradient needed to maintain hydrostatic equilibrium). Such departures are another defining characteristic of galaxy-scale gaseous haloes \citep[see also][]{Oppenheimer2018}, which deviate from the HSE more strongly than the ICM in galaxy cluster haloes \citep[e.g.][]{Lau2013, Nagai2013, Zhuravleva2016}. These deviations appear to be larger than those measured in the CGM simulations analyzed in \citet{Lochhaas2019}, possibly due to the idealized nature of their set-up, which does not account for cosmological accretion and substructure. Nonetheless, we do find that one-dimensional analytic models assuming HSE can provide an approximate description of these haloes (see discussion in the next section).

\begin{figure}
    \centering
    \includegraphics[width=\columnwidth]{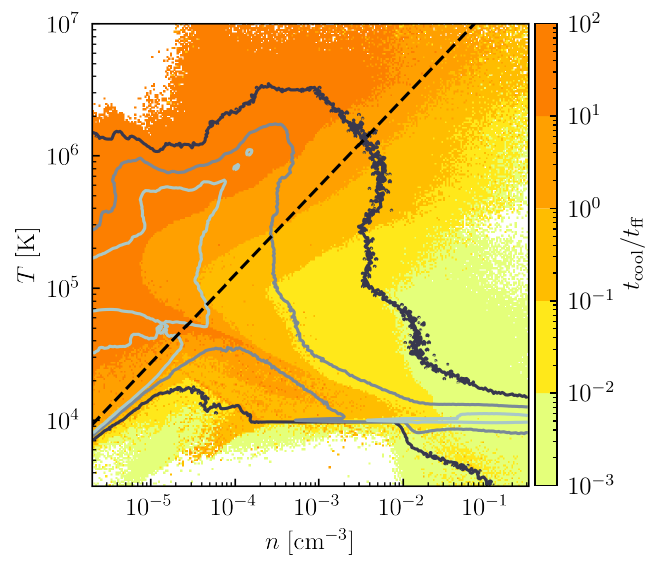}
    \caption{The ``phase diagram'' of number density and temperature for CGM gas in all simulations combined, again at $z=0.25$, colored by its mass-weighted average $\tctff$ (excluding negative values, which we interpret as unshocked IGM accretion).  The grey contours show the mass-weighted PDF of CGM gas, as in Figures~\ref{fig:radpro} and ~\ref{fig:prof_tctff}, with lighter shades showing higher values of the PDF (enclosing $68$, $95$, and $99\%$ of the mass). The dashed black line shows the specific entropy threshold used to separate the ``hot'' and ``cool'' phases, $K = \Klow\equiv5\;{\rm keV}\;{\rm cm}^2$.}
    \label{fig:n_T_tcool_tff}
\end{figure}

Further insight into the general properties of these haloes can be gained from the distribution of thermodynamic quantities and their median profiles as a function of galactocentric radius shown in Figure~\ref{fig:radpro}. Quantities in every panel exhibit large scatter at all radii, indicating the inhomogeneous nature of the CGM in these simulations. We use the median value to define one-dimensional profiles for individual simulated galaxies (shown as colored lines) at each radius, calculated \textit{excluding} substructures and low entropy gas. The median profiles are roughly self-similar but do exhibit some object-to-object variation, which arise due to differences in evolutionary histories of these galaxies (see Figure~\ref{fig:radpro_each} in Appendix \ref{sec:indpro} for the distributions in individual haloes).

The distribution of number density reveals a significant dense component at $n \approx 10^{-2}{\rm cm}^{-3}$. This cool gas resides at the radiative cooling/photo-ionization heating equilibrium of $T \approx 10^{4}{\rm K}$ that is widespread at all radii $r \lesssim 0.5\Rvir$. The narrowness of this ridge in {\it total free particle} number density reflects the rapid change in the ionization state of gas due to self-shielding from the cosmological UV background \citep[see][for details]{FIRE2}. This cool gas co-exists with the volume-filling hot phase that peaks at roughly the virial temperature. These quantities combined give a specific entropy ($K$) distribution that is spread over $\approx 3-4$ orders of magnitude, quantitatively re-enforcing the impression of the entropy map in Figure \ref{fig:maps}: gas with a wide range of entropies co-exist at the same radius in the multiphase CGM produced by these simulations. The median entropy profiles are also quite shallow at all radii ($d\ln K/d\ln r \lesssim 0.5$), suggestive of both cooling and feedback, since both processes tend to flatten the entropy profile \citep[e.g.][]{Sharma2012b}. Median density profiles display a logarithmic slope that tends to $d\ln n/d\ln r = -1$ at small radii and $\sim -2.5$ at larger radii, which can be directly compared to analytical models of the CGM such as those presented in \citet{MillerBregman2013, LiBregman2017}.

The middle-right panel of Figure~\ref{fig:radpro} shows that the bi-modality in the distributions of density, temperature, and entropy are not apparent in the pressure distribution. This indicates that gas of different temperatures is in rough pressure balance, except for the gas in high-pressure spikes associated with substructures. Nevertheless, pressure does vary by more than an order of magnitude at each radius indicating the presence of non-isobaric fluctuations in these haloes. These fluctuations are one of the contributing factors to the  deviations from hydrostatic equilibrium that we noted above. 

The large range ($\gtrsim 2$ dex) of gas metallicities that can be seen in the lower right panel of Fig.~\ref{fig:radpro} and their overall values ($\sim 1$ dex lower than the galaxy ISM) are consistent with some observational estimates of the CGM gas \citep[e.g.,][]{Kacprzak2019}, although perhaps not others \citep[e.g.,][]{Prochaska2017}. A detailed comparison of these simulated gaseous haloes to observations is beyond the scope of this analysis, but will be presented in a forthcoming paper (Hummels et al in prep.). 

The $\tctff$ timescale ratio in the CGM of our simulations as a function of radius is shown in Figure~\ref{fig:prof_tctff}. This quantity exhibits a wide range of values and a bi-modal distribution with typical values of $10^{-3}-0.1$ and $1-100$ for gas in the cool and hot phases, respectively.  Median values in the hot phase are approximately constant at $\tctff\approx 5-20$ throughout the CGM in all simulations. These characteristic values of $\tctff$ are similar to those typically measured for galaxy clusters and consistent with the expectations from the precipitation model. However, we show in Section~\ref{sec:accretion} that there is no particular threshold value of $\tctff$ in the hot phase that predicts gas cooling.

Another perspective on the thermodynamic state of the CGM is shown in Figure~\ref{fig:n_T_tcool_tff}, where the distribution of gas number density and temperature for all five halos combined is shown color-coded by $\tctff$. Grey curves show constant-level contours of the mass-weighted PDF, and the dashed line indicates the entropy threshold we use to define the hot and cool phases. The hot halo occupies a broad distribution in $T\sim 10^5 - 10^6{\rm K}$ and $n \sim 10^{-5} - 10^{-3}{\rm cm}^{-3}$ and the isothermal cool phase is primarily located at $T\sim 10^4\,{\rm K}$ and $n \gtrsim 10^{-2}\,{\rm cm}^{-3}$, consistent with Figure~\ref{fig:radpro}. We interpret the gas at $T\lesssim 10^5{\rm K}$, $n\lesssim 10^{-5}{\rm cm}^{-3}$ as unshocked IGM accretion. The specific entropy we chose to separate the hot and cool phases corresponds to CGM gas with $\tctff \sim 1$ in the hot halo. The distributions for each individual simulation are shown in Figure~\ref{fig:n_T_tcool_tff_each} in Appendix~\ref{sec:indpro}.

\subsection{Comparison to one-dimensional ``precipitation'' models}
\label{sec:modelcomp}

\begin{figure*}
    \centering
    \includegraphics[width=\textwidth]{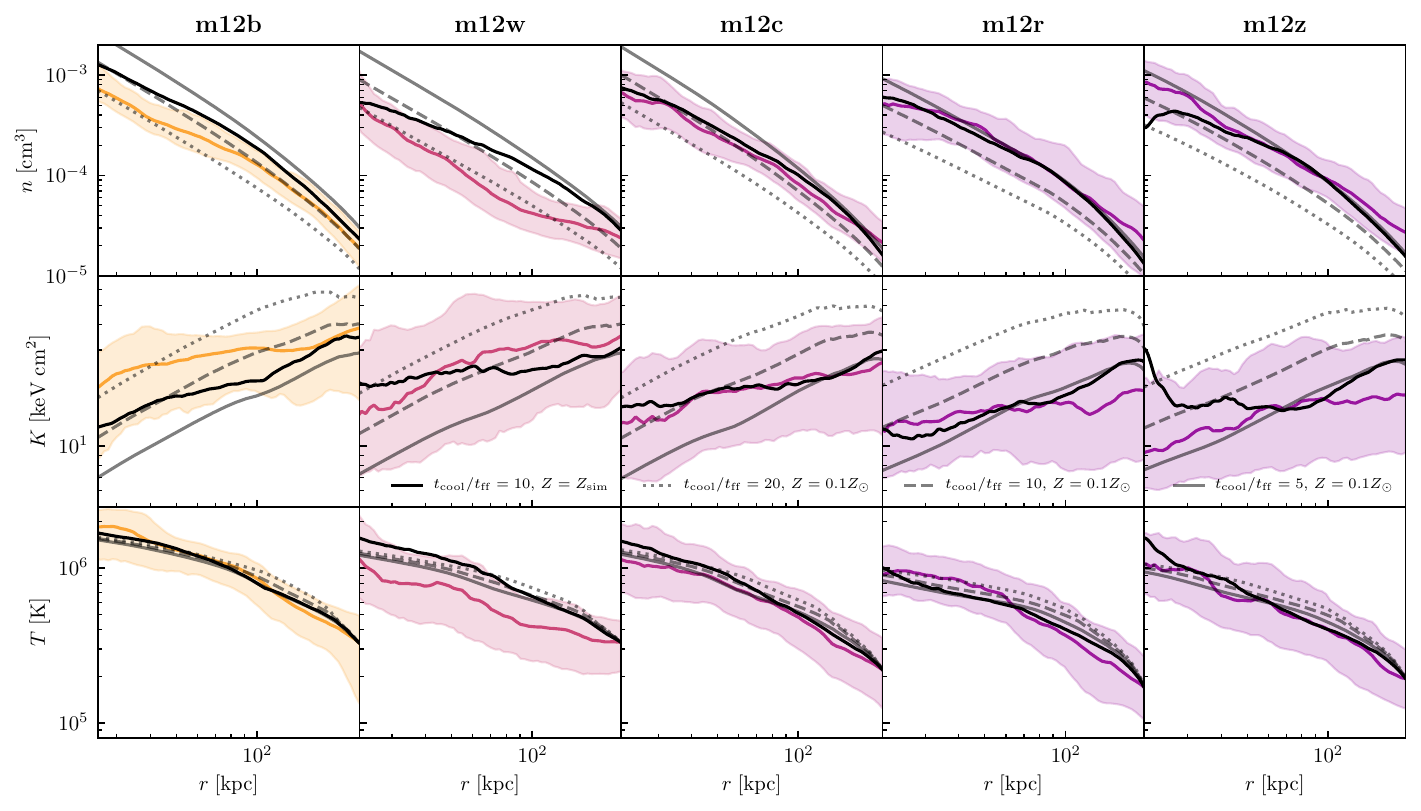}
    \caption{Predictions of the one-dimensional precipitation model compared to simulation profiles. Solid color lines (same colors as in Fig.~\ref{fig:radpro}) with shaded regions show the median gas number density, (upper row) specific entropy (middle row), and temperature (lower row) profiles and the 16th and 84th percentiles of gas particles around them. Black and grey lines of different styles are predictions of the precipitation model described in Section~\ref{sec:modelcomp}. The $\tctff$ threshold, metallicity used to calculate the cooling rate, and boundary condition on temperature (imposed at $\Rvir$) are all essentially free parameters in this model. We adopt the temperature at the virial radius equal to that measured in each simulation for all models shown. We show model predictions for a range of assumed $\tctff$ values that encompass the values found in the simulations (5-20).  The solid black lines show model predictions using the metallicity profile of each simulation and $\tctff = 10$, while gray lines show model predictions with a constant metallicity of $0.1Z_{\odot}$ and $\tctff$ values of 5, 10, and 20 in different line styles.}
    \label{fig:oneD_models}
\end{figure*}

We begin our quantitative comparison to the predictions of CGM thermal instability models by analyzing the high-entropy hot halo of our simulations. Specifically, we compare the predictions of one-dimensional models motivated by the precipitation scenario to the radial profiles of thermodynamics quantities in the hot phase of the simulated CGM. 

In the precipitation ansatz, gas below a constant threshold $\tctff = \xi$ precipitates out of hot CGM. This threshold thus imposes a constraint on the relationship between the thermodynamic properties of the CGM and the gravitational potential. This is typically expressed as a limiting number density profile, obtained by expanding the definition of the cooling time:
\begin{equation}
    \tcool = \frac{n \tau}{(\gamma - 1)\nH^2\Lambda(\tau, Z, \nH)} = \xi \tff
\end{equation}
to give
\begin{equation}\label{equ:nH_precip}
\nH(r) = \frac{\tau(r)}{(\gamma - 1) X(r) \mu \Lambda[\tau(r), Z(r), \nH(r)] \xi t_{\rm ff}(r)}
\end{equation}

\noindent (where $\tau = k_{\rm B} T$, and we have used the identity $\nH = X\mu n$). The values of  $\gamma = 5/3$, $X = 0.75$, and $\mu = 0.6$ can be safely assumed for the hot phase of the CGM, and $\xi$ is a free parameter of the model by design. Additional constraints are required to specify $\tau(r)$, $Z(r)$, and $\tff(r)$. The HSE assumption provides such a constraint by relating $\nH$, $\tau$, and $\tff$ in a first-order ordinary differential equation (ODE)

\begin{equation}\label{equ:tau_ODE_precip}
\frac{d\ln\tau}{d\ln r} = -\left(\frac{d\ln \nH}{d\ln r} + \frac{2\mu\mp r^2 }{t_{\rm ff}^2\tau}\right),
\end{equation}
where we have implicitly assumed that $X$ is constant. We then substitute Eq.~\ref{equ:nH_precip} to obtain a first-order ODE in $\tau$, 
\begin{multline}
\frac{d\ln\tau}{d\ln r} = \left[\left(\frac{\partial\ln\Lambda}{\partial\ln Z}\frac{d\ln Z}{d\ln r} + \frac{d\ln t_{\rm ff}}{d\ln r}\right)\left(1 + \frac{\partial\ln\Lambda}{\partial\ln \nH}\right)^{-1}-\frac{2 \mu \mp r^2}{t^2_{\rm ff}\tau}\right]\\ \times\left[1 + \left(1 - \frac{\partial\ln\Lambda}{\partial\ln\tau}\right)\left(1 + \frac{\partial\ln\Lambda}{\partial\ln \nH}\right)^{-1}\right]^{-1}.
\end{multline}
\noindent which can be numerically integrated given the metallicity profile $Z(r)$, the free-fall timescale profile $\tff(r)$, and a boundary condition on temperature at some radius. This model is similar to the model presented in \citet{Sharma2012b}, where it was compared to data on galaxy clusters.\footnote{We note that our implementation is not identical to that of \citet{Sharma2012b} because of differences in how we determine the boundary condition; \citet{Sharma2012b} do not require the outer parts of their profile to satisfy a constant $\tctff$, since the cooling time of gas at the largest radii is so long that it is unlikely to participate in precipitation. Instead, \citet{Sharma2012b} choose a boundary condition in pressure that ensures a power-law entropy profile in HSE at large radii, and then modify the profiles for radii in the ``core'' where $\tctff$ drops below the precipitation threshold. The simplification of our implementation is purely for computational convenience, and while it may unphysically imply precipitation in gas with extremely long cooling times,  this should be unimportant for the purposes of our analysis, as \citet{Sharma2012b} found that the constant $\tctff$ core extended out to the virial radius for $\Lstar$-mass galaxies in their calculations. Therefore, a more physical implementation would only impact predictions at the outermost radii (if at all), on which our qualitative results do not crucially depend.}

We consider a ``best-case scenario'' for the precipitation model in which we set  $\xi = 10$ (in approximate agreement with the simulations, see Figure~\ref{fig:radpro}), and take $Z(r)$ from the simulation directly. We also show predictions for a range of assumed values for the constant $\xi$ that bracket the values seen in the simulations while assuming a constant $Z(r)=0.1Z_{\odot}$ (also approximately representative of the simulations). In addition, we adopt $\tff(r)$ and the $T(\Rvir)$ boundary condition measured in the simulations. In any practical application of this model to observational data, these quantities would need to be assumed or independently constrained, thereby increasing the uncertainty of the model predictions.

We compare model profiles computed using these assumptions to the median profiles measured in the simulations in Figure~\ref{fig:oneD_models}. The constant metallicity model profiles can generally be brought in agreement with the number density profile of each simulation by fine-tuning the $\tctff$ threshold value, but they all predict entropy profiles steeper than in the simulations. However, when we use metallicity gradients measured in the simulations (see Figure~\ref{fig:radpro}) agreement improves considerably, even if we adopt the same threshold of $\tctff = 10$ for all simulations. This agreement is consistent with the results of Figure~\ref{fig:prof_tctff}, which shows that the CGM of these simulations has an approximately constant median $\tctff$ profile. The most successful models shown predict CGM gas masses similar to those in the simulations, and none exceed the universal baryon fraction. Therefore, self-consistent choices for the model's gravitational potential, temperature boundary condition, $\tctff$ value, and metallicity profile result in fairly accurate (to within a factor of $\sim 2$) predictions of the number density and entropy profiles in the simulations.

The model of \citet{Sharma2012b} has been extended to both the steadily growing observational data on massive group and cluster systems from X-ray emission \citep{Voit2015, Voit.Donahue2015} and the rapidly emerging absorption line data on the CGM of less massive galaxies \citep{ Voit2019a, Voit2019b}. The version presented in the latter papers has also been used to model scaling relations in galaxy properties across the entire galaxy mass range \citep[see also][]{Voit2015Galaxies}. However, we note that the model described in these papers is {\it not} identical to the one introduced in \citet{Sharma2012b} or presented in our analysis. 

Specifically, the version of the model presented in \citet{Voit2019a} combines the number density profile Eq.~\ref{equ:nH_precip} and an assumption that the shape and normalization of $T(r)$ are set by the halo circular velocity profile to compute a ``precipitation-limited'' entropy profile. This entropy profile is then held fixed and combined with the HSE equation to obtain other thermodynamic quantities as a function of radius. As we show in the Appendix~\ref{sec:model_appendix}, this version predicts entropy profiles that are significantly steeper than in the simulations, even when adopting the simulation metallicity profile. This is mainly because the temperature profile assumed in the \citet{Voit2019a} model has shape and normalization significantly different from $T(r)$ in the simulated halos. The model of \citet{Sharma2012b} fares better, because it does not assume a temperature profile, but rather self-consistently combines the assumptions of a constant $\tctff$ and HSE to predict thermodynamic profiles.

Finally, we emphasize that the precipitation model of \citet{Sharma2012b} is successful  only in a very approximate sense, i.e. within a factor of $\sim 2$ in the predicted quantities. This is because the model assumptions, particularly HSE, are only roughly realized in the simulations. In principle, one could extend the model to account for additional sources of pressure -- such as turbulence, rotation, and infall, as has been done in other models of the CGM by \citet{Faerman2017, Faerman2019} -- which might correct for these discrepancies, but we consider this outside the goals of this analysis.

\subsection{Origin of low entropy gas}
\label{sec:lowKorigins}
While a simple model based on the assumption of a constant $\tctff$ time and approximate hydrostatic equilibrium is reasonably successful in describing the median profiles of the high-entropy CGM gas in our simulations, this comparison at a single snapshot alone does not validate the dynamical process of thermal instability as the origin of cool circumgalactic gas. However, given the fully time-dependent and 3D information provided by the simulations, this basic physical process can be tested directly by analyzing the thermodynamic history of tracer particles. As we noted before, the gaseous haloes in the FIRE-2 simulations \textit{are} manifestly multiphase. In this section we therefore examine the origin of the cool phase in these simulated CGM and the physical processes driving gas cooling.

\begin{figure}
\centering
\includegraphics[width=\columnwidth]{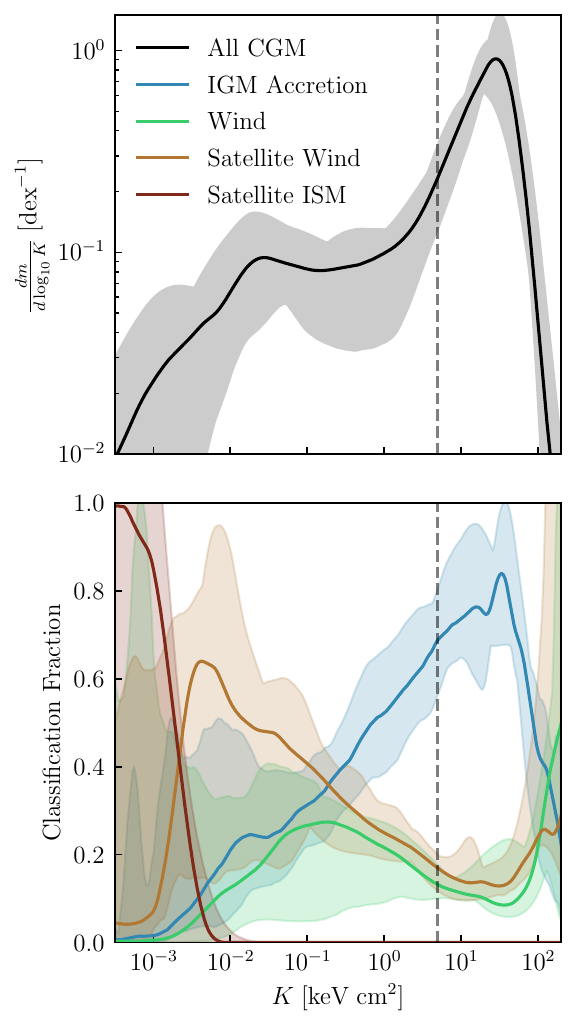}
\caption{Specific entropy PDF of the CGM.
    \textit{Top panel:} mass-weighted PDF of specific entropy for the CGM. The solid line indicates the distribution of CGM gas of all simulations combined at $z = 0.25$, while shaded bands indicate the full range for individual simulations. This distribution is dominated by an approximately log-normal component that peaks near $K_{\rm vir}\sim 10-20 \;{\rm keV}\;{\rm cm}^2$, and a tail below $\Klow \equiv 5\;{\rm keV}\;{\rm cm}^2$, extending to lower specific entropy. This tail constitutes $\sim20-50\%$ of the mass in the CGM.
    \textit{Bottom panel:} fraction of mass in each tracking classification \citep[described in][]{Hafen2019} as a function of specific entropy. Red indicates the mass fraction classified as satellite ISM, orange ``wind'' from a satellite galaxy, green ``wind'' from the central galaxy, and blue accretion from the IGM. The hot phase is dominated by particles classified as IGM accretion, whereas the low-entropy tail is dominated by satellite wind, with substantial contributions from all other categories, and large halo-to-halo scatter.} 
\label{fig:K_pdf}
\end{figure}

\begin{figure*}
\includegraphics[width=0.7\textwidth]{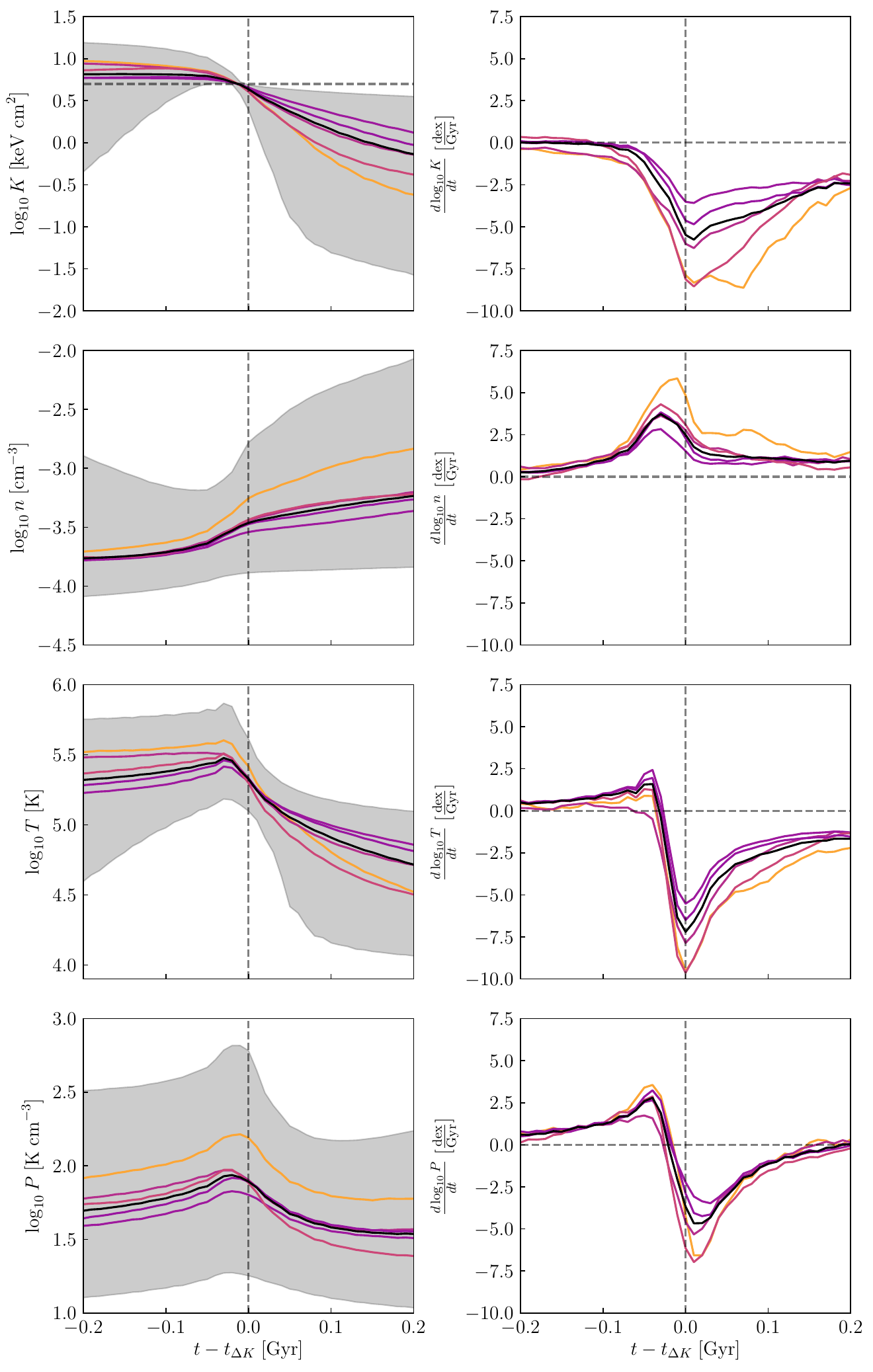}
\caption{Thermodynamic histories of cooling CGM gas. For each low-entropy gas particle in the CGM at $z=0.25$, we identify the time $t_{\Delta K}$ corresponding to the \textit{most recent snapshot at which the particle cooled from the hot halo}. This is defined as the most recent snapshot after crossing $K = \Klow\equiv 5\;{\rm keV}\;{\rm cm}^2$ (shown as a horizontal dashed line in the top left panel). For all panels in this figure, we use the thermodynamic history of each particle \textit{as a function of the time difference from} $t_{\Delta K}$. Thus, zero on the x-axis corresponds to $t_{\Delta K}$ for every particle used to calculate the displayed quantities.
    \textit{Left column:} From top to bottom, the entropy, number density, temperature, and pressure as a function of the time difference from $t_{\Delta K}$ for each particle. Solid lines show the median history for each halo (colored lines, using the same color-to-simulation mapping as in previous figures),and all haloes combined (black line). The shaded region encompasses the 16th and 84th percentiles at each time for all halos combined.
    \textit{Right column:} Derivatives of median histories. Solid lines show time derivatives of the median histories displayed in the left column smoothed by a top hat filter of width $30{\rm Myr}$. Dashed grey lines indicate 0 on the $x$ and $y$ axes for visual aid. The gas density increases \textit{before} the temperature and specific entropy decrease, \textit{indicating that cooling of low entropy gas from the hot halo is primarily the result of density perturbations}. As well, note that the pressure of the gas initially {\it increases}, unlike the isobaric or isochoric perturbations typically assumed for linear thermal instability.}
\label{fig:tracks}
\end{figure*}

\begin{figure*}
    \centering
    \includegraphics[width=0.7\textwidth]{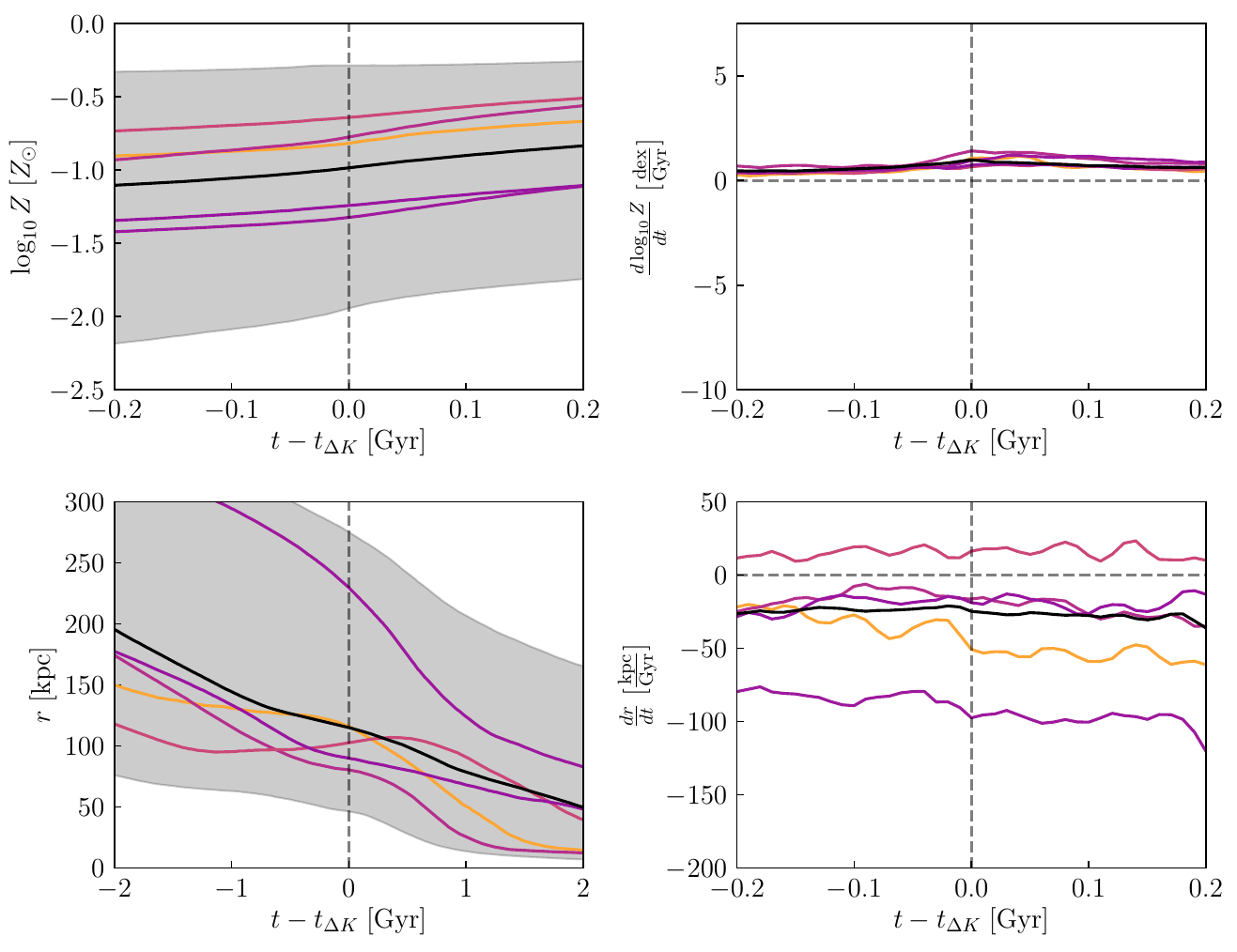}
    \caption{Same as Fig.\ref{fig:tracks}, but for the metallicity $Z$ and radius $r$ of gas tracers. We show an extended time-frame for the tracks in radius to highlight the lack of uplift in the cooling CGM gas within the past few Gyr. Gas cooling does not appear to be triggered by or immediately precede either significant changes in metallicity or radius for CGM gas. Simulation {\bf m12r} appears as an outlier in the radial tracks because most of the cool gas in this halo is associated with a large merging subhalo - see Figures \ref{fig:maps},~\ref{fig:radpro_each}.} 
    \label{fig:tracks_pt2}
\end{figure*}

\subsubsection{PDFs of specific entropy}

Figure~\ref{fig:K_pdf} shows the mass-weighted PDF of specific entropy for the CGM gas in our simulation sample at $z=0.25$ (top panel) and the proportional contribution of each particle tracking classification determined in \citet[][bottom panel]{Hafen2019b}.\footnote{The PDFs are computed using a Kernel Density Estimate (KDE) with the Epanechnikov kernel of constant size set to the optimal value using Silverman's ``rule-of-thumb'' \citep{Silverman1986}.} Solid lines show the mean PDFs for all 5 simulations combined, while the shaded regions show the full range of individual simulation distributions. These PDFs were calculated using only the gas tracers for which the tracking analysis was performed, which constitute a representative subset of the full CGM.

The specific entropy distribution at $z=0.25$ consists of two main components: the mass-dominant hot halo with an approximately log-normal distribution centered at roughly the virial entropy of these haloes ($\sim 10-20 \,{\rm keV}\;{\rm cm}^2$) and the low-entropy gas with a wide distribution below $\sim 5\,{\rm keV}\;{\rm cm}^2$. We therefore adopt the value of $K = \Klow\equiv 5 \;{\rm keV}\;{\rm cm}^2$ as the boundary between low and high entropy phases of the CGM. Most ($\gtrsim 70\%$, see column 6 of Table~\ref{table:simulations}) of the low-entropy phase by this definition was previously in the high-entropy phase. This indicates that most of the low-entropy CGM gas in these simulations is the result of cooling from the hot halo.

The fraction of CGM gas in each ``origin'' categorization as defined in \citet{Hafen2019}, as a function of entropy, is also informative. This is presented in the bottom panel of Figure~\ref{fig:K_pdf}, showing that the hot phase originates primarily from the shock-heated gas accreted from the IGM.  At lower entropy, the fraction of gas accreted from the IGM rapidly decreases, but nevertheless remains a dominant component as low as $\approx 0.3\,\rm keV\,cm^2$. At even lower values, the low entropy gas is dominated by the ``satellite wind'' classification, with roughly equal contributions to the remaining gas from central galaxy wind and IGM accretion. This strongly suggests that gas expelled from satellite galaxies via feedback-driven winds or tidal debris either remains at low entropy or must cool rapidly as it is ejected. In practice we see evidence for both in the simulations, to be discussed further in Section~\ref{sec:therm_hist}.

We note that at low entropy values $K\le\Klow$, the relative fractions of low-$K$ gas due to different tracer classes vary significantly between individual halos. This reflects the complicated, dynamic nature of the CGM in these simulations discussed in Section~\ref{sec:genprop} and highlights multiple possible mechanisms whereby low-entropy gas is produced, the relative contributions of which vary depending on the evolutionary history of an individual halo. The overall picture that emerges from this analysis is that the CGM in these simulated galaxies is a highly dynamic environment.

\subsubsection{Thermodynamic history of low-entropy gas: cooling triggered by wide-spread, non-linear perturbations}\label{sec:therm_hist}

To better quantify the processes that produce low-entropy gas, we analyze the thermodynamic histories of tracers that have cooled from the high to low entropy phase by $z=0.25$. Specifically, for each gas tracer $i$ with $K_i(z=0.25)< \Klow$, we identify the time $t_{\Delta K}$ corresponding to the closest prior snapshot before which the tracer had entropy $>\Klow$, i.e. we select the snapshot immediately after the tracer crosses $K=\Klow$ from above for the last time before $z=0.25$. By this definition, we exclude from this portion of the analysis gas which remained at an entropy below this threshold for the entire simulation prior. We also require that the gas tracers have $n < 0.13\;{\rm cm}^{-3}$ when they cool to avoid contamination from dense gas that is cooling in the ISM of satellite galaxies. This allows us to examine trends in thermodynamic quantities around the time $t_{\Delta K}$ at which cooling in the CGM is actually occurring.

Figure~\ref{fig:tracks} shows $K$, $n$, and $T$, and $P$ and their derivatives as a function of time around $t_{\Delta K}$ for the tracers. Colored solid lines show the median values of the quantities at every time for each halo. The black line shows the median and the shaded regions in the left column show the 16th and 84th percentiles for tracers from all simulations. The time derivatives of the individual simulation medians are shown in right column. These derivatives were smoothed by a top hat filter of width $30\,{\rm Myr}$, slightly larger than the time between snapshots. This figure shows that the most rapid decrease in entropy and temperature, representing the most rapid cooling, occurs slightly {\it after} $t_{\Delta K}$, while gas number density increases most rapidly slightly {\it before}\/ $t_{\rm \Delta K}$. 

These evolutionary patterns indicate that cooling typically starts with compression on a time scale of $\approx 0.05-0.1$ Gyr. The compression is nearly isothermal, with only a mild increase of temperature due to adiabatic heating. The magnitude of density perturbations is typically large (factor of $\sim 1.5-2$) and compression thus significantly decreases the cooling time, which scales approximately as $1/n$. The compressed gas then cools rapidly while its density continues to increase. During the peak of $dn/dt$ gas pressure also increases, in contrast to the isobaric or isochoric perturbations usually assumed for linear thermal instability \citep[e.g.,][]{Field1965, McCourt2012}. Pressure then sharply declines as rapid cooling begins.

The primary driver of gas cooling thus appears to be rapid and large-amplitude compressive perturbations. Since these trends can be identified in each individual simulation and the combination thereof, they appear to be general properties of the perturbations that drive cooling in the CGM of $\Lstar$ galaxies simulated with the FIRE-2 model.

Figure~\ref{fig:tracks_pt2} is a similar evolutionary plot for tracer galactocentric radius $r$ and metallicity $Z$, neither of which appears to be a dominant cause of cooling. In all but one halo, these tracers show small radial velocities consistent with the pressure-supported high-entropy halo (see the lowest left panel of Figure~\ref{fig:radpro}) prior to cooling. Cooling does not appear to dramatically change the trajectories of CGM particles, in contrast to the expectation of simple precipitation-motivated models for galactic accretion (see Section~\ref{sec:accretion} and Figure~\ref{fig:accretion} for further discussion).

Furthermore, note that the vast majority of the tracers begin at large radii and subsequently accrete upon cooling. This suggests that the ``condensation due to uplift'' process emphasized by previous investigations \citep[e.g.]{Li.Bryan2014, Voit2017} as a possible trigger of thermal instability in the ICM is a subdominant process in the CGM of these simulated galaxies. However, we emphasize that this conclusion may strongly depend on both the implementation of feedback in these specific simulations, as well as the cosmic epoch we analyze, since uplift-induced thermal instability may well operate in Milky Way-mass galaxies that are driving stronger winds into the CGM. 

Figure~\ref{fig:tracks_pt2} also shows that metallicity increases rapidly only {\it after} the onset of cooling. This implies that the enhancement of the cooling rate due to chemical enrichment of the CGM gas is not a primary trigger of the cooling we identify. The metallicity increase after the onset of cooling is likely a result of gas mixing during the advection of condensing gas to lower radii, where metallicity tends to be larger (see Figure~\ref{fig:radpro}).

\begin{figure}
    \centering
    \includegraphics[width=\columnwidth]{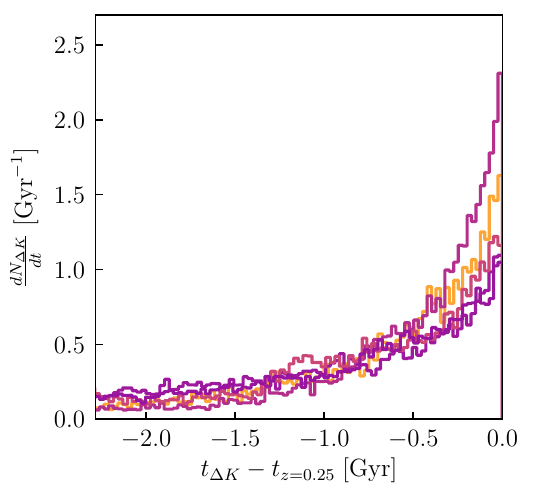}
    \caption{Distribution of $t_{\Delta K}$, the times at which gas tracers undergo cooling, as a function of look-back time from $t_{z = 0.25}$. Different colors represent different simulated haloes using the same color-to-simulation mapping as in previous figures. The distributions are quite similar, despite markedly different accretion histories of individual haloes in our sample over this time period.}
    \label{fig:tdelta_dist}
\end{figure}

\begin{figure}
    \centering
    \includegraphics[width=\columnwidth]{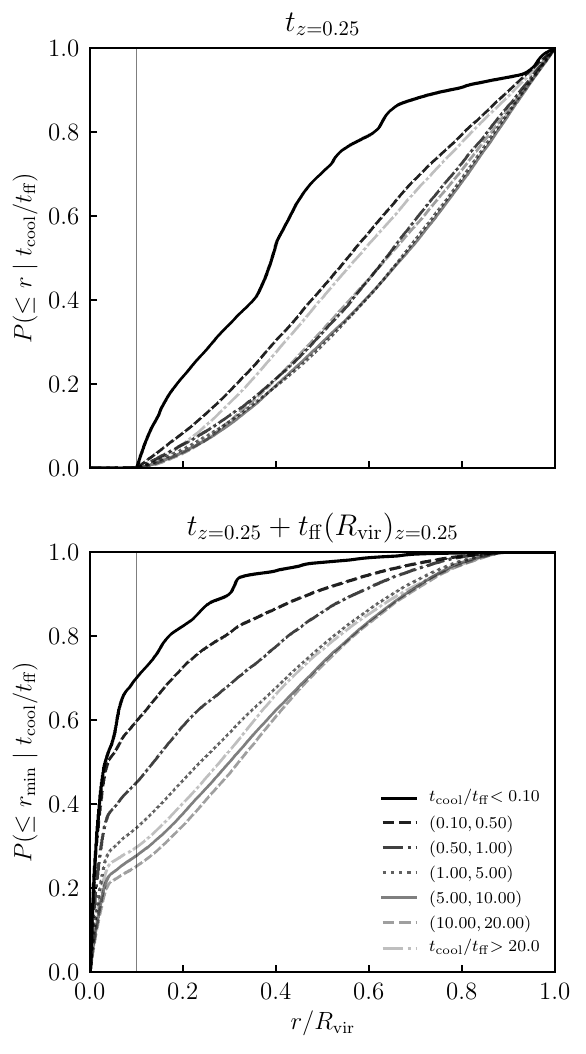}
    \caption{Accretion histories of CGM gas as a function of initial $t_{\rm cool}/t_{\rm ff}$. 
    \textit{Top panel:} CDFs of CGM gas cells as a function of galactocentric radius at $z=0.25$ binned by $\tctff$ at that snapshot. These distributions are calculated combining all simulations in our sample. Increasingly transparent colors show CDFs of increasing initial $\tctff$ value as indicated in the legend on the bottom panel. The narrow vertical line in both panels indicates $0.1\Rvir$, approximately corresponding to the edge of the central galaxy.
    \textit{Bottom panel:} CDFs of the minimum radius attained by each resolution element in the top panel one free-fall time (at $\Rvir$, $\approx 2.3$ Gyr) later. Note that the lower-panel CDFs are calculated including star particles that formed from gas particles represented in the top panel, as well as those that remained ISM or CGM gas. Only $\tctff$ values below 1 indicate a significantly greater likelyhood of accretion onto the central galaxy.}
    \label{fig:accretion}
\end{figure}

The onset of cooling does not appear to be sensitive to the recent galaxy accretion history or star formation rate. Figure~\ref{fig:tdelta_dist} shows that the distribution of $t_{\Delta K}$ in the past 2.3 Gyr (the free-fall time at the virial radius for these halos) for individual tracers is very similar for all simulations, despite marked differences in the properties of these simulated galaxies at this time. For example, {\bf m12r} and {\bf m12w} were chosen to have an LMC-mass satellite at low redshift, and are therefore undergoing a significant merging events. Thus the processes that lead to cooling are generic and operate continuously, implying that the perturbations that trigger this cooling are ubiquitous and constantly generated by the dynamic processes that influence these haloes.

We note that for all the simulations we analyze, Figure~\ref{fig:tdelta_dist} shows that the amount of gas that has cooled appears to strongly increase as a function of time. We interpret this as resulting from a lack of significant heating in the CGM of these simulations at the late cosmic epochs we analyze. Indeed, \citet{Stern2021} find that the halos in these simulations are well-described by cooling flows. We therefore expect most of the cool gas at any one time to have cooled recently.

\subsection{Gas accretion from the CGM}
\label{sec:accretion}

The precipitation model of the CGM on galactic scales is interesting both because of its potential to explain the observation of ubiquitous multiphase gas around massive galaxies (probed by absorption lines in quasar spectra) and because of the role that precipitation may play in regulating gas accretion onto galaxies and thus their star formation histories. In the previous section we demonstrated that the CGM of Milky Way-mass haloes do undergo widespread cooling as a result of primarily non-linear density perturbations. Here we explore the degree to which the $\tctff$ value of a given parcel of gas predicts this cooling.

Figure~\ref{fig:accretion} shows the radial distribution of gas tracers in all simulations binned by {\it initial} (a.k.a. at $z=0.25$) $\tctff$  at $z=0.25$ (top panel) and one $\tff\approx2.3\;{\rm Gyr}$ later (bottom panel). If a constant threshold $\xi$ in $\tctff$ determined whether a given gas parcel precipitated out of the CGM and accreted onto the central galaxy, we would expect the radial distributions of these components to be markedly different at a later time $t\gtrsim \tff$, with gas particles of small $\tctff<\xi$ concentrating at small radii and gas with $\tctff>\xi$ still in the CGM close to their original distribution. Indeed, gas particles at initial $\tctff$ values $\lesssim 1$ reach significantly lower minimum radii within the following $2.3$ Gyr, indicating that they are more likely to accrete onto the central galaxy than gas particles at higher initial $\tctff$. However, Figures~\ref{fig:radpro}, \ref{fig:prof_tctff}, and \ref{fig:n_T_tcool_tff} show that this gas is almost entirely composed of the low-entropy ``cool'' phase, indicating that it had already cooled from the hot halo. This gas therefore sinks through the ambient hot halo to lower radii because it is already overdense and is no longer buoyant. This is consistent with the analysis of CGM ``fates'' in \citet{Hafen2019b}, where it was shown that cool gas at $T\sim10^4\,{\rm K}$  predominantly accretes onto the central or satellite galaxies (see the top left panel of their Figure 6). 

However, for gas particles initially in the hot halo with $\tctff > 1$ the distributions in the bottom panel of Figure~\ref{fig:accretion} are similar, showing that there is no correlation between initial the $\tctff$ value and the subsequent change in radius of these gas parcels. This strongly suggests that $\tctff$ {\it does not predict subsequent cooling from the hot halo or accretion onto the central galaxy}. A similar analysis (not shown) indicates that the initial radial velocity of the CGM gas is also a relatively poor predictor of subsequent accretion onto the central galaxy. This is likely because random or ``turbulent'' motions are comparable in magnitude to bulk radial flows in these simulations. 

We note that while the $\approx 10-20\%$ of the hot halo gas that accretes onto the central galaxy is {\it proportionally} much less than the $\gtrsim 40\%$ of gas from the cold phase that accretes, since most of the {\it total} CGM gas mass is in the hot phase, the accretion rate onto the central galaxy will be dominated by gas that was initially in the hot halo. We also note that additional analysis reveals that the very low initial $\tctff$ tracers that remain at large radii are strongly radially clustered, suggesting association with satellite galaxies.

\section{Discussion}
\label{sec:discussion}

We begin our discussion with the definition of the term ``thermal instability'' to avoid ambiguity. Broadly speaking, a parcel of gas is said to undergo thermal instability if it is perturbed so that its entropy evolves away from the entropy of the surrounding gas instead of returning to its initial value, resulting in co-existence of different gas phases \citep{Field1965, Balbus1986}. We note that this definition makes no reference to the scale or amplitude of the perturbation, nor to the thermodynamic state of the unperturbed background. Consequently, any runaway cooling that establishes a persistent contrast in entropy with ambient gas is encompassed by this definition of thermal instability. In addition, large ($\delta \gtrsim 1$) perturbations can still result in thermal instability even if small-amplitude perturbations are stable according to the linear stability analysis. 

Maps (Figure~\ref{fig:maps}) and radial profiles (Figure~\ref{fig:radpro}) indicate the presence of ubiquitous density perturbations in the simulated CGM. A Lagrangian tracking analysis (Figure~\ref{fig:tracks}) demonstrates that these density perturbations lead to widespread cooling of CGM gas from the $\sim 10^5-10^6\,{\rm K}$ and $n\lesssim10^{-3}{\rm cm}^{-3}$ hot halo into dense, clumpy and filamentary cool gas with $T\sim10^4-5\times10^4\,{\rm K}$. The spatial distribution of this cool gas (as shown in Figure~\ref{fig:maps}) and its origins (Figure~\ref{fig:K_pdf}) strongly suggest that these perturbations are sourced by the cosmological accretion of gas from the IGM, stellar feedback-driven outflows (including from satellite galaxies), tidal interactions, and ram pressure stripping. Many aspects of this process are qualitatively consistent with the precipitation scenario, in which the multiphase nature of the observed CGM around galaxies is the result of thermal instability.  However, the low-entropy phase in the FIRE-2 simulations does not form from small, isobaric seed perturbations evolving into small clouds, which is often envisioned in linear stability analyses  \citep[e.g.,][]{McCourt2012,Choudhury_Sharma2016}. Instead, the CGM in the simulated haloes is subjected to density perturbations with a wide range of amplitudes (Figure~\ref{fig:tracks}) and spatial scales (Figures~\ref{fig:maps},~\ref{fig:radpro}) many of which are beyond the domain of validity encompassed by linear stability analyses.  

In the remainder of this section we discuss implications of our results for ``precipitation'' models of the CGM and interpretation of simulation results. We also discuss comparisons to previous studies in the literature and the potential effects of simulation resolution on our results. 

\subsection{Interpretation of results and comparisons with previous studies}
\label{sec:interp_comparison}

We find that the low-redshift CGM of $\Lstar$ galaxies in the FIRE-2 simulations is multiphase and dynamic, with significant ($20-50\%$) deviations from hydrostatic equilibrium. The phase structure of these halos can be broadly characterized as bimodal in temperature and entropy, composed of a hot halo which is volume filling and a cool phase which has clumpy and filamentary morphology. Temperatures in the hot halo are near or slightly less than the virial temperature $T\sim 10^5-10^6\,{\rm K}$ while the cool phase is mostly in photo-ionization heating/cooling equilibrium at $T=10^4\,{\rm K}$. All thermodynamic quantities show an orders-of-magnitude range {\it at all radii}, indicating the inhomogeneous nature of the CGM in these simulations. 
Nevertheless, we find that the median radial profiles of thermodynamic quantities of the hot gas vary smoothly. In particular, the specific entropy $K$ and cooling-to-free-fall time ratio $\tctff$ in these simulations are approximately constant with radius.

Consequently, we show that a simple one-dimensional model of the hot CGM based on \citet{Sharma2012b} which assumes a constant $\tctff \approx 10$ value and hydrostatic equilibrium can provide a reasonable description for the simulated profiles {\it if and only if} we use the metallicity profile, $\tff(r)$ profile, and temperature at the virial radius measured in simulation in calculating the predictions of this model. Moreover, we re-iterate that our results demonstrate $\tctff$ to be a poor predictor of gas cooling and accretion from the hot halo (see Fig.~\ref{fig:accretion}), in contradiction to the assumptions of precipitation models \cite[e.g.][]{Sharma2012b, Voit2015Galaxies}. 

Interestingly, the characteristic values of $\tctff$ we find in the FIRE-2 haloes are close to the predictions of the cooling flow model and simulation results of \citet{Stern2019}, which exhibit shallow profiles of $\tctff$ with characteristic values of $\tctff\sim 1-10$ for objects of this halo mass. The $\tctff\sim 1-10$ are also close to the prediction of the CGM model of \citet{Faerman2019} that assumes a constant entropy profile, although we note that this model additionally attempts to incorporate the effects of magnetic field and cosmic ray pressure support, which are neglected in the simulations we analyze. These results further demonstrate that shallow radial profiles of $\tctff$  can be generated by the mechanisms other than local thermal instability. 

As stated at the end of Section~\ref{sec:modelcomp},  the model of \citet[][]{Sharma2012b} -- a variant of which was used in our analysis -- differs  from the ``precipitation limited'' model of \citet{Voit2019a}.  In fact, these two models can make qualitatively different predictions in some regimes. While the \citet{Voit2019a} model also begins with the constraint on the number density profile imposed by the assumption of a constant $\tctff$ value (Eq.~\ref{equ:nH_precip}), the model presented in \citet{Voit2019a} fixes the entropy profile in the ``precipitation-limited'' regime to $K_{\rm pre}(r)\propto [\Lambda(T,Z)r]^{2/3}$ by assuming that $k_B T(r)=\mu m_p v_c^2(r)$ where $v_c(r)$ is the halo circular velocity profile.  The $K_{\rm pre}(r)$ profile with this assumption is then used to solve the HSE equation (see Eq.~\ref{eq:Kpre} and the associated discussion in Appendix~\ref{sec:model_appendix}). However, this assumed temperature profile turns out to be significantly different in shape and amplitude from the actual $T(r)$ of gas in the FIRE simulations. Consequently, the entropy profiles of the \citet{Voit2019a} model are significantly steeper than those in the simulations (see Figure~\ref{fig:oneD_models_Voit}). We also note that this formulation of the model does not guarantee a constant $\tctff$ with radius. 

Regardless of the differences between {\it median} radial profiles predicted by different models of the CGM, our analyses also indicate the potentially crucial role played by the {\it scatter} in thermodynamic properties at a given radius. If the CGM of real $\Lstar$ galaxies is as inhomogeneous as predicted by the FIRE-2 simulations, then ability of simple models to match observational data will require correctly predicting the full distribution of thermodynamic quantities at a given radius, not just a single characteristic value. This is especially true for the comparison to absorption-line observations, which are sensitive to thermodynamic structure on arbitrarily small scales. Indeed, even the minimalist models of \citet{Faerman2017}, \citet{Faerman2019}, and \citet{Voit2019a} adopt a PDF of thermodynamic quantities at a given radius for precisely these reasons.

Indeed, we explicitly show that the $\tctff$ values of {\it hot} gas of the simulated haloes  in the FIRE-2 simulations span a wide range of $\sim 1-100$ at all radii. This is qualitatively consistent with the idealized simulations of the CGM presented in \citet{Fielding2017}, who find that the hot CGM exhibits wide variations of $\tctff$ (specifically in their simulation with winds using mass loading factor of $\eta=0.3$, see their Fig. 13), which they also interpret as evidence for a thermal instability driven precipitation/feedback cycle. A large range of $\tctff$ was also found in the idealized simulations of \citet{Choudhury2019}, who showed that the minimum $\tctff$ required for cooling and condensation of gas out of the hot phase monotonically increases with increasing amplitude of seed gas density perturbations, indicating that it is easier for perturbations of larger amplitude to condense \citep[see also][]{Pizzolato.Soker2005,Singh.Sharma2015,Meece2015}. This dependence is particularly strong for perturbations of $\delta \gtrsim 1$, such as those in the CGM of FIRE-2 simulations. Note, however, that simulations of \citet{Choudhury2019} imposed net thermal equilibrium using a heating term in the entropy equation set to the average value of radiative cooling at a given radius, while the FIRE-2 haloes may lack such a global thermal balance. It is therefore not clear whether the ${\rm min}(\tctff)$ thresholds derived by \citet{Choudhury2019} are applicable to the CGM in the FIRE-2 simulations.

More generally, our analysis suggests that that cosmological accretion and satellite galaxies contribute substantially in generating density perturbations and shaping the dynamics and phase structure of the CGM gas in our simulations. Idealized simulations that attempt to model the {\it global} dynamics of the CGM, but do not include these processes will thus be limited in their validity. Our results thus {\it motivate} the full exploration of environments and physical processes relevant to CGM physics on all scales, through both self-consistent fully cosmological {\it and} carefully designed idealized numerical simulations. 

In the ICM regime, where precipitation has been explored thoroughly, idealized simulations may not be subject to the same limitations. Observations show that intracluster plasma in such cluster cores 1) is characterized by a relatively steep and ``universal'' entropy profile \citep[e.g.,][]{Cavagnolo2009,Babyk2018}, 2) has a narrow range of relatively small density perturbations \citep[e.g.,][]{Zhuravleva2018}, 3) is close to hydrostatic equilibrium in most cases (i.e. maximum deviations of $\sim10-20\%$). These are the physical conditions under which the {\it linear} thermal instability driven precipitation is expected to operate \citep{McCourt2012, Sharma2012, Sharma2012b} and for which simulations do predict a well-defined threshold in $\tctff$ for thermal instability \citep[e.g.][]{Li2015}. This is likely why simple analytic models based on the constant $\tctff$ assumptions provide a good match to the properties of the ICM in simulated and observed groups and clusters \citep[e.g.,][]{Sharma2012b}, although we note that fully cosmological simulations of galaxy clusters have not yet been applied to this problem. 

Interestingly, there is evidence for thermal instability stimulated by the uplift of hot gas from, e.g., AGN feedback in the ICM of cluster-sized systems \citep[e.g.,][]{Li.Bryan2014}. As  emphasized by \citet{Voit2017}, the approximately adiabatic uplift of low entropy gas to larger radii can promote thermal instability in two ways: 1) the decreased ambient pressure as a function of radius means that uplifted gas will cool due to expansion, potentially to temperatures at which the cooling function is greatest and 2) the increased ambient entropy removes the stabilizing effect of buoyancy \citep{Balbus.Soker1989}.  However, Figure~\ref{fig:tracks_pt2} shows that little, if any, of the cooling gas in our simulations has experienced recent uplift. This is consistent with the general lack of outflows that reach significantly into the CGM from galaxies of this mass at late cosmological times \citep{Muratov2015, AnglesAlcazar2017}. 

The qualitatively different properties of circumgalactic gaseous halos in the $\Lstar$ regime revealed by our analysis imply that the application of this physical model to the entire range of galaxy masses well below the group and cluster scale \citep[e.g.]{Voit2015} may not be warranted.  

During the last stages of this paper's finalization, \citet{Nelson2020} appeared on the arXiv. This interesting study also explores the origin of cool gas around massive galaxies in $M_{\rm vir}\sim 10^{13}\, M_{\odot}$ dark matter haloes. Their analysis thus focuses on galaxies more massive than ours and uses simulations of lower resolution that employ a different hydrodynamic solver and an entirely different set of prescriptions for the physical processes that regulate galaxy formation. Nevertheless, they conclude that thermal instability due to non-linear density perturbations sourced in part by cosmological substructures are the primary drivers of cooling in the CGM of their simulated haloes, encouragingly similar to our findings.  

\subsection{Effects of resolution and neglected physical processes}
\label{sec:resolution}

\begin{table}
    \centering
    \resizebox{\columnwidth}{!}{\begin{tabular}{c|c|c|c|c|c|c|c|}
    \hline
    \hline
    &  & $m_{\rm b} = 7100M_{\odot}$ & & & & $m_{\rm b} = 57000M_{\odot}$ \\
    \hline
    Name & $M_{\rm CGM}$ & $f_{{\rm low}\;K}$ & $h_{\rm b}$ & & $M_{\rm CGM}$ & $f_{{\rm low}\;K}$ & $h_{\rm b}$ \\
    & [$10^{10}M_{\odot}$] & & [kpc] & & [$10^{10}M_{\odot}$] & & [kpc]\\
    \hline
    {\bf m12b} &  5.23 & 0.18 & 3.50 & & 5.13 & 0.06 & 6.76 \\
    {\bf m12c} &  3.04 & 0.23 & 3.62 & & 3.67 & 0.10 & 6.92\\
    {\bf m12r} & 3.44 & 0.46 & 2.75 & & 3.86 & 0.44 & 6.09 \\
    {\bf m12w} & 2.80 & 0.38 & 4.16 & & 3.71 & 0.13 & 7.27 \\
    \end{tabular}}
    \caption{Effect of resolution (quantified by $m_b$, the MFM gas cell mass) on CGM mass $M_{\rm CGM}$, low-entropy mass fraction $f_{{\rm low},K}$, and median MFM gas cell kernel size $h_{\rm b}$. $M_{\rm CGM}$ and $f_{{\rm low},K}$ are computed as defined in Table~\ref{table:simulations}.}
    \label{tab:f_low_res_test}
\end{table}

Recent extremely high-resolution, idealized simulations of small volumes with initial conditions motivated by the CGM \citep{McCourt2018, Liang2019} have suggested a characteristic scale for multiphase gas clouds set by the product of the sound speed and the cooling time, which is of order $0.1-100$ pc for $10^4$ K gas. This is several orders-of-magnitude less than the typical gas cell kernel size in the CGM of our simulations, which thus certainly lack the spatial resolution needed to capture the shredding and fragmentation of gas to structures of these scales. Nevertheless, our analysis shows that many of the perturbations responsible for widespread cooling in our simulations are large-scale and therefore certainly resolved.

\begin{figure*}
    \centering
    \includegraphics[width=\textwidth]{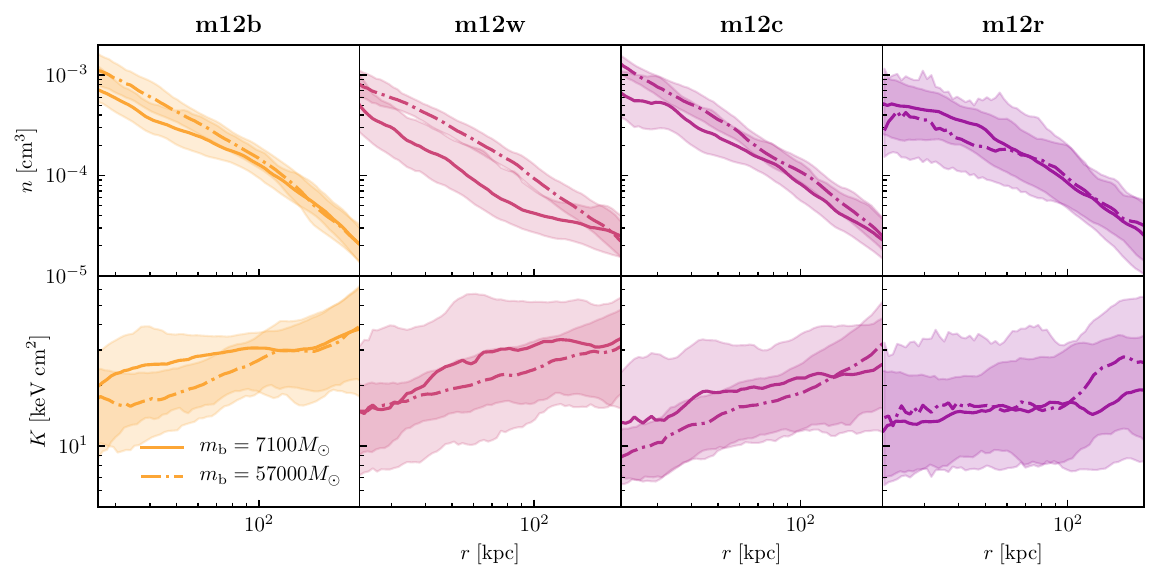}
    \caption{Test of resolution dependence. Median radial profiles of number density (top row) and specific entropy (bottom row) of gas for four of the five simulations in our sample. Solid lines are the profiles measured in the simulations used in this study (shown in Figure \ref{fig:radpro}), while the dot-dashed lines are the same halos re-simulated with $8\times$ more massive cells in the zoom-in region, as indicated in the legend. The effect of resolution does not appear to be systematic or larger than the halo-to-halo or stochastic snapshot-to-snapshot variation.}
    \label{fig:radpro_res_test}
\end{figure*}

To assess the resolution dependence of our results, in Figure~\ref{fig:radpro_res_test} we compare the distribution of the number density and specific entropy in the CGM of the {\bf m12b,c,r,w} haloes re-simulated at a roughly order-of-magnitude lower resolution (gas cells of $m_b = 57000 \Msun$ compared to $m_b = 7100 \Msun$) to the profiles in the simulations used in this study. The figure shows substantial differences in the profiles at two resolutions, but the differences do not appear to be systematic. Moreover, the differences are smaller than the halo-to-halo variations. Although this does not prove that the results would not change qualitatively if resolution was increased by orders of magnitude, this test shows that our results that involve median radial profiles of the hot phase are likely robust to modest changes in resolution. 

We also report the total CGM mass, mass fraction of low entropy gas, and median CGM gas cell kernel size  in Table~\ref{tab:f_low_res_test} for these lower resolution simulations. Here, we do see a systematic trend with resolution: the lower resolution simulations appear to have much less cool CGM gas. This suggests that the cooling processes responsible for the formation of this cold gas may not be converged in the simulations we analyze. However, we note that the changes in the total CGM mass are also significant, and one halo's ({\bf m12r}) cold gas mass fraction does not change significantly with resolution, suggesting that some of this difference may be due to stochastic variation in the properties of the CGM as a function of time as a consequence of the chaotic nature of galaxy formation, and not necessarily to unconverged results.

This is consistent with the findings of several recent investigations that presented simulations with significantly enhanced resolution in the CGM gas of zoom-in cosmological simulations of $\Lstar$ galaxies \citep[e.g.][]{vandeVoort2019, Peeples2019, Hummels2019}. These studies also showed weak resolution dependence of the properties of the hot phase of the CGM, but stronger dependence of the mass and spatial distribution of the cool phase.

A final caveat to this study is the importance of physical processes neglected by these simulations. First of all, there is ongoing debate about effects of specific implementations of star formation and stellar feedback processes that are well-established fundamental drivers of galaxy evolution at this mass range. For example, the inclusion of cosmic ray feedback in cosmological galaxy formation simulations has been found to have a potentially profound effect on the dynamic properties of the CGM around $\Lstar$ galaxies \citep[e.g.][]{Ji2019}. At the same time, the relative importance of AGN feedback in the haloes of this mass is not yet fully understood. Furthermore, recent studies have shown that the presence of magnetic fields \citep[e.g.][]{Ji2018, Wang2019b} and plasma physics processes such as cosmic rays and conduction \citep[e.g.][]{Li2019, Kempski2019} can have important effects on thermal instability and the survival of multiphase gas in hot haloes. Moreover, there is some indication that the default FIRE-2 prescriptions with which the simulations we analyze were run may fail to quench star formation of galaxies at any mass \citep{Su2019}, and produce $\Lstar$ galaxies with systematically higher gas fractions and star formation rates than observed. This suggests too much gas flows from the CGM to the central galaxy in these simulations. The resolution of this tension could in principle produce a CGM closer to hydrostatic equilibrium than we find in the current generation of simulations, in which case linear thermal instabilities that give rise to a well-defined $\tctff$ threshold could play a larger role. 

For the reasons above, it is by no means certain that the physics of the CGM is captured correctly in the FIRE-2 simulations we analyze in this study. However, these simulations are quite successful at reproducing observed galaxy scaling relationships \citep[e.g.,][]{Ma2016, FIRE2} and $\Lstar$ galaxy satellite galaxy properties \citep[e.g.,][]{Wetzel2016, Garrison-Kimmel2019, Samuel2020}. They are also consistent with many existing observational constraints on the properties of the the CGM itself \citep{FaucherGiguere2015, FaucherGiguere2016, vandeVoort2016, Hafen2017, Ji2019}. We therefore employ this model as an appropriate first exploration of the circumgalactic precipitation picture. As such, this study should be viewed as an initial test of the precipitation scenario rather than the final judgement upon it. It also provides motivation for future tests of this scenario in both idealized and zoom-in cosmological simulations that include a wider range of physical processes, such as those discussed above. 

\section{Summary and conclusions}

In this study we have examined the thermodynamic state and cooling in the low-$z$ circumgalactic medium of cosmological zoom-in FIRE-2 galaxy formation simulations of Milky Way-mass haloes with the specific aim of understanding the origin of its multiphase structure. Our results and conclusions are as follows: 

\begin{itemize}
    \item[(i)] We find that the CGM in the simulations is generally multiphase and dynamic, exhibiting significant bulk flows and deviations from hydrostatic equilibrium (Section~\ref{sec:genprop} and Figures~\ref{fig:maps}, \ref{fig:radpro}). The phase structure of the CGM in these simulations is defined by a high-entropy ``hot halo'' at $T\sim 10^5-10^6\,{\rm K}$ and a cool phase at $T\sim10^4\,{\rm K}$. The hot halo is volume-filling, dominates the mass budget, and exhibits a wide range  of $\tctff$ values, $\tctff\sim 1-100$, with median values $\sim 5-20$. The cool gas constitutes $\sim20-50\%$ of the CGM mass in different halos and is distributed in clumps, filaments, and satellite-associated substructure. 
    \item[(ii)] We show that a one-dimensional model constructed  assuming a constant $\tctff \approx 10$ and hydrostatic equilibrium is able to match the median radial density and entropy profiles in simulations reasonably well (i.e. to within a factor of $\sim 2$), provided that we use free-fall time, gas metallicity profiles, and temperature at the virial radius measured in the simulations when calculating the predictions of this model (Section~\ref{sec:modelcomp}). However, the practical application of this model to observations will require assumptions about these quantities, increasing the uncertainty in these predictions. Specifically, we show that when the metallicity profile is assumed to be constant (unlike in the simulations), the model predicts entropy profiles that are significantly steeper than the actual simulated profiles. 
    \item[(iii)] We also show that the ``precipitation-limited'' model of \citet{Voit2019b} predicts entropy profiles inconsistent with those measured in the simulations, even if the simulation metallicity and $\tctff$ profiles are used. This is because the temperature profile assumed in this model differs significantly from the simulation temperature profiles in both shape and normalization (see Section~\ref{sec:modelcomp} and Appendix~\ref{sec:model_appendix} for further discussion).
    \item[(iv)] We investigate the origin of the multiphase structure of the CGM in the FIRE-2 simulations with a particle tracking analysis and find that most ($\gtrsim 70\%$) of the low-entropy gas in these haloes has cooled from the hot halo (Table~\ref{table:simulations}, Section~\ref{sec:lowKorigins}). This cooling is a result of thermal instability triggered by large amplitude density perturbations. We find evidence (Figure~\ref{fig:K_pdf}) that these perturbations are sourced by cosmological accretion, feedback driven winds and the debris of tidal interactions from the central and satellite galaxies. They are generally non-isobaric, and many are large enough that linear thermal instability theory likely fails to describe their evolution (Figure~\ref{fig:tracks}). Adiabatic uplift of low-entropy gas and changes in the cooling rate due to chemical enrichment both appear to be of secondary importance to density fluctuations in the production of cool gas from thermal instability in these halos (Figure~\ref{fig:tracks_pt2}).
    \item[(v)] The wide distribution of density perturbation amplitudes likely results in a wide range of conditions conducive to cooling, which correspond to a range of $\tctff$ thresholds for thermal instability instead of a single value. Haloes in these Milky Way-scale simulations are therefore qualitatively different from the approximately hydrostatic and comparatively smooth ICM of galaxy clusters. 
    \item[(vi)] We explicitly show that gas in the ``cool'' phase is more likely to accrete onto the central galaxy, but the $\tctff$ value of hot gas parcels does {\it not} predict whether this gas will cool in the first place (Figure~\ref{fig:accretion}). Our results therefore suggest that the extrapolation of a $\tctff$ threshold in thermal instability from the ICM of galaxy clusters to lower mass halos may not be warranted.
\end{itemize}

Our findings reveal the complex, multiphase, and non-equilibrium physical processes characteristic of the CGM of Milky Way-mass galaxies in a suite of state-of-the-art galaxy formation simulations. While the physical realism of the gaseous halos in {\it any} galaxy formation simulation remains difficult to assess because of limitations inherent to both the current observational data and numerical models, our results strongly motivate further theoretical investigation of these processes in realistic cosmological settings.

\section*{Acknowledgements}
We thank Mark Voit for clarifying feedback on our comparisons to his work, Hsiao-Wen Chen for providing references to relevant observational results, and the anonymous referee for a careful and thorough report. CE and AK were supported by the NSF grants AST-1714658 and AST-1911111.
CAFG and ZH were supported by NSF through grants AST-1517491, AST-1715216, and CAREER award AST-1652522, by NASA through grant 17-ATP17-0067, by STScI through grants HST-GO-14681.011, HST-GO-14268.022-A, and HST-AR-14293.001-A, and by a Cottrell Scholar Award from the Research Corporation for Science Advancement.
EQ was supported in part by a Simons Investigator Award from the Simons Foundation and by NSF grant AST- 1715070.
JS is supported by the CIERA Postdoctoral Fellowship Program.
DK was supported by NSF grant AST-1715101 and the Cottrell Scholar Award from the Research Corporation for Science Advancement.
AW received support from NASA through ATP grant 80NSSC18K1097 and HST grants GO-14734, AR-15057, AR-15809, and GO-15902 from STScI; a Scialog Award from the Heising-Simons Foundation; and a Hellman Fellowship.

Our analysis made use of the following publicly available software packages: Matplotlib \citep{Matplotlib}, SciPy \citep{SciPy}, NumPy \citep{NumPy}, and COLOSSUS \citep{COLOSSUS}.

\section*{Data availability}

The data supporting the plots within this article are available on reasonable request to the corresponding author. A public version of the GIZMO code is available at \url{http://www.tapir.caltech.edu/~phopkins/Site/GIZMO.html}. Additional data including simulation snapshots, initial conditions, and derived data products are available at \url{https://fire.northwestern.edu/data/}.



\bibliographystyle{mnras}
\bibliography{bibliography} 


\appendix

\section{Comparison to the precipitation model of Voit et al. 2019}
\label{sec:model_appendix}

\begin{figure*}
    \centering
    \includegraphics[width=\textwidth]{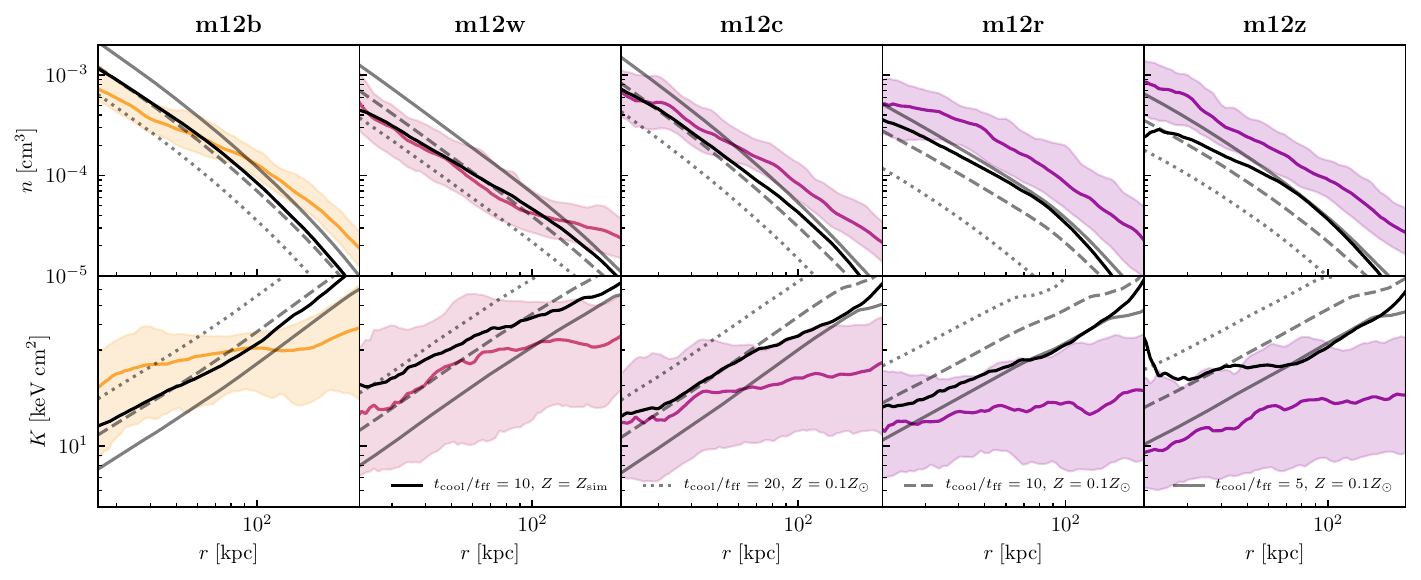}
    \caption{One-dimensional ``Precipitation-Limited'' models from \citet{Voit2019a} compared to simulation profiles. Colored lines show number density of all particle species (upper row), and specific entropy (lower) profiles with shading encompasing the 16th and 84th percentiles for the simulated hot haloes, as shown with the same colors in Figure~\ref{fig:radpro}, while black and grey lines of different styles are predictions of the \citet{Voit2019a} precipitation model assuming the same $\tctff$ values and metallicity profiles as in Figure~\ref{fig:oneD_models}. As with the implementation explored in the main text, the predicted number density profile can be tuned to match those of each simulation by varying the assumed $\tctff$ and metallicity, but the entropy profiles are generally steeper than those in the simulations, even when the metallicity profiles are taken from the simulations themselves.}
    \label{fig:oneD_models_Voit}
\end{figure*}

As noted in Section~\ref{sec:modelcomp} and discussed in Section~\ref{sec:interp_comparison}, the one dimensional precipitation model for CGM profiles presented in this analysis is different from the ``precipitation-limited'' model of \citet{Voit2019a} and its earlier variants. In this Appendix, we compare the two versions of the model and repeat the comparison to simulation profiles with the model described in \citet{Voit2019a}. While we refer readers to that paper for a complete discussion of their model and its comparison to observations, we summarize the main features of the model and differences from with \citet{Sharma2012b} below. 

The \citet{Voit2019a} model also assumes that gas reaches a constant $\tctff$ precipitation threshold in the central regions where precipitation regulates gas properties. As shown in Section~\ref{sec:modelcomp}, this implies that the radial number density profile \citep[expressed in electron number density for consistency with the expressions shown in][]{Voit2019a} is related to the temperature profile, $T(r)$ in the following way:
\begin{equation}\label{eq:npre}
n_{\rm e}(r) = \frac{(1 - \gamma)k_{\rm B} T(r)}{\xi\tilde{\Lambda}[T(r),Z(r),\nH(r)]}\,\frac{n}{n_{\rm i}}\,\frac{v_{\rm c}(r)}{\sqrt{2}}\, r^{-1},
\end{equation}
where $k_{\rm B}$ is the Boltzmann constant, $Z(r)$ is the gas metallicity profile, $n$ is total gas number density, $n_i$ is number density of ions, $v_{\rm c}(r)=[GM(<r)/r]^{1/2}$ is the circular velocity determined by the gravitational potential, $\gamma = 5/3$ is the adiabatic index for an ionized plasma, $\xi$ is the assumed limiting value of $\tctff$, and $\tilde{\Lambda}$ is the radiative cooling function 
\citep[which calculate using the tables from][consistent with the cooling tables used in the FIRE-2 source code, see Section~\ref{sec:radcool}]{Wiersma2009}. $\tilde{\Lambda}$ is a function of the temperature $T(r)$, and metallicity $Z(r)$, and total hydrogen (neutral and ionized) number density $\nH(r)$ profiles. Since the \citet{Wiersma2009} tables define the cooling function $\Lambda$ such that $\nH^2\Lambda$ is the net radiative cooling rate, we define $\tilde{\Lambda} = \nH^2\Lambda/(n_e n_i)$ to distinguish between the different cooling function definition used in the derivations of \citet{Voit2019a}, in which the net volumetric cooling rate is $n_e n_i\tilde{\Lambda}$. 

As discussed in Section~\ref{sec:interp_comparison}, the density condition Eq.~\ref{eq:npre} can be combined with the HSE condition to give an ODE in temperature which can be numerically integrated, as we do in the model used in our paper. Instead, \citet{Voit2019a} uses the density profile of Eq.~\ref{eq:npre} with additional assumptions about the temperature profile to motivate an entropy profile $\Ke(r)$ which is then combined with the HSE equation to give a {\it different} ODE in temperature. 

Specifically, the model of \citet{Voit2019a} adopts an entropy profile of the form 
\begin{equation}
\Ke(r) = k_{\rm B}T n_{\rm e}^{-2/3} = \Kbase(r) + \Kpre(r)
\label{eq:Ke}
\end{equation}
where
\begin{equation}
\Kbase(r) = K_{\mathrm e,0} \left(\frac{r}{\R200c}\right)^{1.1}
\end{equation}
is the self-similar ``baseline'' specific entropy profile obtained from simulations of structure formation with non-radiative gas dynamics \citep{Voit2005} and unaffected by precipitation. $\Kpre(r)$ is the ``precipitation-limited'' entropy profile calculated using Eq.~\ref{eq:npre} and a temperature profile defined by the circular velocity of the gravitational potential $k_{\rm B} T(r) = \mu \mp v_c^2(r)$. With this assumption the ``precipitation-limited'' entropy profile is therefore
\begin{equation}\label{eq:Kpre}
\Kpre(r) = (2\mu m_{\rm p})^{1/3}\left[\frac{n}{n_i}\frac{(\gamma - 1)}{\xi}\right]^{-2/3}\tilde{\Lambda}^{2/3}[Z(r), T(r),\nH(r)]\,r^{2/3}.
\end{equation}
We use the subscript $e$ to clarify that the definition of the specific entropy adopted in \citet{Voit2019a} uses electron density, while total gas number density of all free particles (including electrons) is used in the definition of entropy in the model used in our paper. Here $\mu=0.6$ is the assumed mean molecular weight, $m_{\rm p}$ is the mass of the proton, and $n_i/n=0.44$ is the ionic-to-total number density ratio. Note that even with a constant metallicity profile $Z(r)$, $K_{\rm e,pre}$ will not generally scale as $r^{2/3}$ because of the temperature (and thus radial) dependence of the cooling function.  

With the additional constraint provided by this entropy profile, the HSE equation can be expressed as 
\begin{equation}
\frac{d\ln\tau}{d\ln r} = \frac{3}{5}\frac{d\ln \Ke}{d\ln r} -\frac{4\mu \mp r^2}{5t_{\rm ff}^2\tau}
\end{equation}
which can be integrated numerically, given an appropriate boundary condition and assumed $\tff$ profile. This formulation therefore imposes the precipitation condition $\tctff=\xi$ through the {\it entropy} profile, which requires the assumption of an initial temperature profile that is not generally guaranteed to be the same as the one predicted by the model. Consequently, the $\tctff$ profile predicted by this model {\it generally depends on radius}, indicating that this model is not a self-consistent implementation of the precipitation ansatz.

We note that the calculations in \citet{Voit2019a} used the cooling function from \citet{SutherlandDopita1993}, which assumes all chemical elements to be in collisional ionization equilibrium, in which case the cooling function can be expressed without a density dependence altogether. Consequently, the model presented in that paper did not need to assume an initial $\nH$ profile to calculate the cooling rate in Eq.~\ref{eq:Kpre}. In order to use a more accurate density-dependent cooling function, we initially calculate the model assuming a constant value $\bar{n}_{{\rm H},200} = 200Xf_b\rho_{\rm cr}/\mp$ with $X=0.75$, $f_b = 0.16$ and $H_0 = 70{\rm km}{\rm s}^{-1}{\rm Mpc}^{-1}$ for $\rho_{\rm cr}$. However, since this is not consistent with the density profile ultimately predicted by the model, we iteratively re-compute the model with cooling rates calculated assuming the density profile predicted by the previous iteration. Iteration is stopped when the density profile is converged (to within a percent fractional difference at all radii). In practice this only requires a few iterations, and the predictions made assuming the initial $\bar{n}_{{\rm H},200}$ are practically indistinguishable from the converged solution. 

Figure~\ref{fig:oneD_models_Voit} repeats the comparison to simulation profiles in Section~\ref{sec:modelcomp} with the model of \citet{Voit2019a}. Here we again use the boundary condition in temperature $T(\Rvir)$ and gravitational potential parameterized by the free-fall time $\tff(r)$ from the simulations. We compare models with the same values of $\xi$ and metallicity profiles as Section~\ref{sec:modelcomp}. The figure shows that the \citet{Voit2019a} version of the precipitation model generally does not match the density and entropy profiles measured in simulations. In particular, the entropy profiles calculated by this model are significantly steeper than those in the FIRE-2 simulations. 

This is a consequence of the assumption imposed in the derivation of Eq.~\ref{eq:Kpre} that $T(r) = \mu \mp v_c^2(r)/k_{\rm B}$, which is generally incorrect for the simulations we analyze. In the simulations, the temperature profiles are significantly lower (by factors of $\sim 1.5-5$) than $\mu \mp v_c^2/k_{\rm B}$ at all radii, and (more importantly) the ratio $T/v_c^2$ is not constant with radius (varying by a factor of $\sim 2-4$ for all simulations over the radial range of our comparison). Consequently, this discrepancy in entropy profile slopes remains for any assumption of the metallicity profile, free-fall time profile, temperature boundary condition, or numerical value of the $\tctff$ threshold. 

\section{Individual Simulation Profiles}
\label{sec:indpro}
\begin{figure*}
    \centering
    \includegraphics[width=\textwidth]{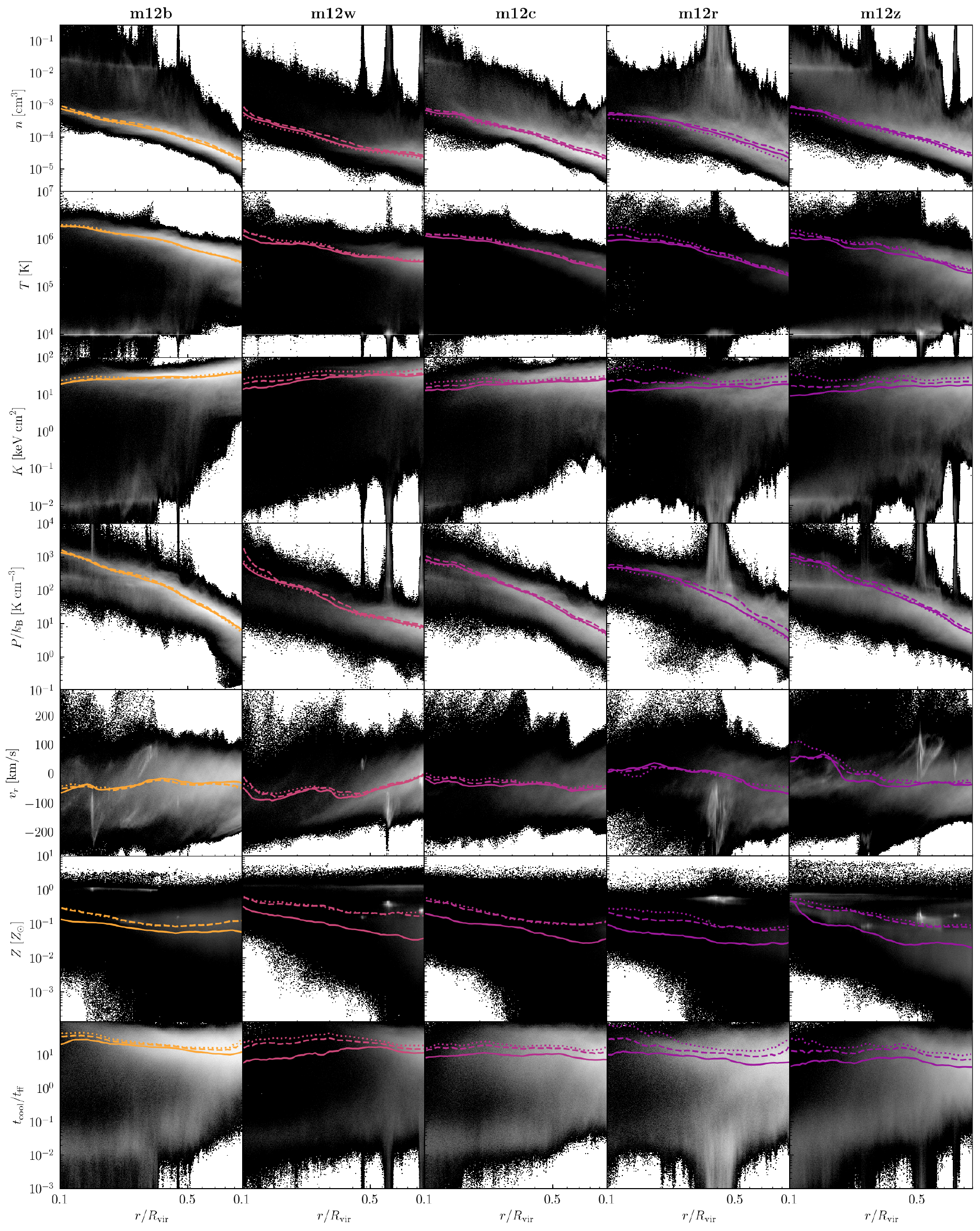}
    \caption{Individual simulation profiles, as shown combined in Figure~\ref{fig:radpro} and ~\ref{fig:prof_tctff}. Solid lines show median profiles of the high-entropy gas (excluding substructure) as a function of radius, while dashed show the mass-weighted mean, and dotted the volume-weighted mean. As in Table~\ref{table:simulations}, the $\tctff$ mean profiles are calculated excluding the small amount of high-entropy gas gas with negative cooling times. The gray scale shows the mass-weighted PDF of all CGM gas (including low entropy gas and substructures).}
    \label{fig:radpro_each}
\end{figure*}

In Figure~\ref{fig:radpro_each} we present the full radial distributions of the quantities shown in Figure~\ref{fig:radpro} for each simulation individually, to highlight the differences potentially masked by their combination. Additionally, we show mass and volume-weighted mean profiles as a function of radius, to demonstrate the (lack of) sensitivity of radial profiles to the statistic used to calculate them. Values of these quantities at characteristic radii are quoted in Table~\ref{table:averaging}. The $\tctff$ mean profiles are calculated excluding the small amount of high-entropy gas with negative cooling times, which we interpret as IGM accretion that has not yet been shock heated. We see that the tail of high density, low temperature, and low entropy gas is a general feature of all simulation distributions, but the amount of mass and its radial concentration vary significantly from halo-to-halo at a given cosmological time.

\begin{figure*}
    \centering
    \includegraphics[width=\textwidth]{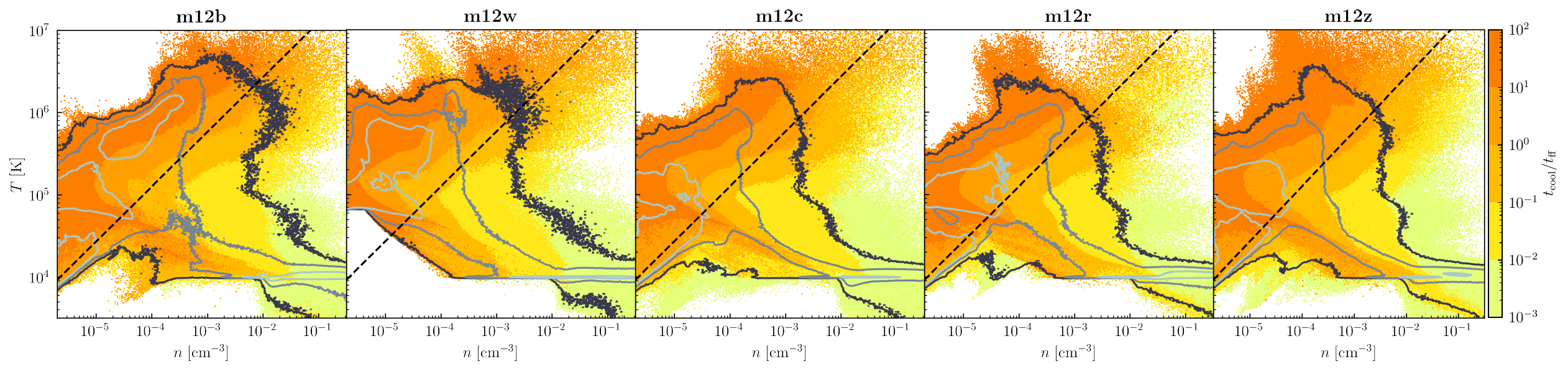}
    \caption{Same as Figure~\ref{fig:n_T_tcool_tff}, but for each simulation separately.}
    \label{fig:n_T_tcool_tff_each}
\end{figure*}

Similarly, Figure~\ref{fig:n_T_tcool_tff_each} shows the distribution of CGM gas in the $n$-$T$ domain for each individual simulation, as shown combined in Figure~\ref{fig:n_T_tcool_tff}.

\begin{table*}
\caption{Effect of averaging on $\tctff$. $t_{\rm cool}$, $\tctff$: median value of the cooling timescale and cooling-to-free-fall timescale ratio in the high-entropy gas, reported at $0.2\Rvir$ and $\Rvir$, as given in Table~\ref{table:simulations}. $\langle t_{\rm cool}\rangle$, $\langle\tctff\rangle$: mass-weighted mean. $\langle t_{\rm cool}\rangle_V$, $\langle\tctff\rangle_V$: volume-weighted mean, calculated by assuming the volume of an individual fluid resolution element scales inversely with its density. We note that in calculating mass and volume-weighted averages we exclude the small amount of high-entropy gas with negative cooling times (due to net photo-heating).}
\begin{tabular}{ccccccc}
\hline
\hline
Name & $\left( \frac{t_{\rm cool}}{t_{\rm ff}}\right)_{0.2R_{\rm vir}}$ & $\left\langle \frac{t_{\rm cool}}{t_{\rm ff}}\right\rangle_{0.2R_{\rm vir}}$ & $\left\langle \frac{t_{\rm cool}}{t_{\rm ff}}\right\rangle_{V, 0.2R_{\rm vir}}$ & $\left(\frac{t_{\rm cool}}{t_{\rm ff}}\right)_{R_{\rm vir}}$ & $\left\langle \frac{t_{\rm cool}}{t_{\rm ff}}\right\rangle_{R_{\rm vir}}$ & $\left\langle \frac{t_{\rm cool}}{t_{\rm ff}}\right\rangle_{V, R_{\rm vir}}$\\
\hline
{\bf m12b}  & 21.2 & 25.1 & 29.1 & 12.9 & 22.3 & 29.9\\
{\bf m12c}  & 11.3 & 18.8 & 24.3 & 7.0 & 11.6 & 16.8 \\
{\bf m12r}  & 11.2 & 17.3 & 43.8 & 6.0 & 23.5 & 31.1 \\
{\bf m12w}  & 9.6 & 23.3 & 32.7 & 12.2 & 15.2 & 19.4 \\
{\bf m12z}  & 5.9 & 11.4 & 16.3 & 4.4 & 7.7 & 11.2\\
\label{table:averaging}
\end{tabular}
\end{table*}

\bsp	
\label{lastpage}
\end{document}